\documentclass[12pt,fleqn]{article}
\usepackage{a4wide,amsmath,axodraw,epsfig}
\begin{document}

\newcommand{\nl}{\notag\\}
\newcommand{\eqn}[1]{Eq.~\!(\ref{#1})}
\newcommand{\fig}[1]{Fig.~\!\ref{#1}}
\newcommand{\suml}{\sum\limits}
\newcommand{\prol}{\prod\limits}
\newcommand{\intinfi}{\int\limits_{-i\infty}^{+i\infty}}
\newcommand{\sumM}{\sum_{n=1}^{M}}
\newcommand{\sumN}{\sum_{k=1}^{N}}
\newcommand{\sumL}{\sum_{\mu=1}^{L}}
\newcommand{\sumn}{\sum\limits_n}
\newcommand{\dsumM}{\sum_{n,m=1}^{M}}
\newcommand{\dsumN}{\sum_{k,l=1}^{N}}
\newcommand{\prodM}{\prod_{n=1}^{M}}
\newcommand{\al}{\alpha}
\newcommand{\ga}{\gamma}
\newcommand{\Ga}{\Gamma}
\newcommand{\ep}{\epsilon}
\newcommand{\la}{\lambda}
\newcommand{\si}{\sigma}
\newcommand{\Si}{\Sigma}
\newcommand{\om}{\omega}
\newcommand{\de}{\delta}
\newcommand{\vDe}{\varDelta}
\newcommand{\vLa}{\varLambda}
\newcommand{\vTh}{\varTheta}
\newcommand{\vt}{\vartheta}
\newcommand{\vp}{\varepsilon}
\newcommand{\ra}{\rightarrow}
\newcommand{\da}{\downarrow}
\newcommand{\pr}{\prime}
\newcommand{\prop}{\mathcal{G}}
\newcommand{\twop}{\mathcal{C}}
\newcommand{\distwop}{\mathcal{B}}
\newcommand{\Dpath}{\mathcal{D}}
\newcommand{\Real}{{\bf R}}
\newcommand{\Comp}{{\bf C}}
\newcommand{\Kube}{{\bf K}}
\newcommand{\intk}{\int_\Kube}
\newcommand{\intkk}{\int_{\Kube^2}}
\newcommand{\rmin}{\textrm{min}}
\newcommand{\half}{{\textstyle\frac{1}{2}}}
\newcommand{\sfrac}[2]{{\textstyle\frac{#1}{#2}}}
\newcommand{\Ord}{{\mathcal O}}
\newcommand{\Dp}{d\mu[\phi]}
\newcommand{\Tr}[1]{\textrm{Tr}\left(#1\right)}
\newcommand{\Exp}[1]{\mathsf{E}\!\left[#1\right]}
\newcommand{\inttext}[3]{\vspace{-#1\baselineskip}\\#3
                         \vspace{-#2\baselineskip}}

\title{{\bf Quantum field theory for discrepancies}}

\author{
Andr\'{e} van Hameren\thanks{andrevh@sci.kun.nl}~ and 
Ronald Kleiss\thanks{kleiss@sci.kun.nl}\\
University of Nijmegen, Nijmegen, the Netherlands}

\maketitle

\begin{abstract}
The concept of {\em discrepancy} plays an important r\^ole in the study of 
uniformity properties of point sets. For sets of random points, the discrepancy 
is a random variable. We apply techniques from quantum field theory to 
translate the problem of calculating the probability density of (quadratic) 
discrepancies into that of evaluating certain path integrals. Both their 
perturbative and non-perturbative properties are discussed.
\end{abstract}

\thispagestyle{empty}

\newpage
\pagestyle{plain}
\setcounter{page}{1}
\newpage

\section{Introduction}
An important actor in numerical integration is the set of points that is used. 
An important factor in the accuracy of the numerical result is the distribution 
of these points in the integration region.
The relatively slow convergence with the number of points of the 
Monte Carlo method for multivariate integration has inspired a search for 
point sets that result in a faster convergence than with sets of random points.
Numerical integration with this kind of point sets goes under the name of 
Quasi Monte Carlo \cite{nieder1}. Whether a point set is 
suitable depends, of course, on the function to be integrated. Therefore, 
it has proven to be useful to assume that the integrand belongs to a certain 
class of functions, the {\em problem class}, 
of which it is a `typical' member. The problem is then translated 
into that of an {\em average-case complexity}, 
the squared integration error made by the numerical integration with the given 
point set, averaged over the problem class \cite{woz,pas,kleiss}. 

If there is not enough information about the 
function to identify it as a member of a problem class, there is no choice 
but to look for point sets that are as uniformly distributed over the 
integration region as possible. In this analysis, measures of non-uniformity 
of point sets, called {\em discrepancies}, are needed, and have been the topic 
of a great number of publications \cite{nieder1,prepubs}. 
A class of these discrepancies, the so 
called {\em quadratic} discrepancies, appear to be identifiable as the 
average-case complexities mentioned before. These also have been the 
inspiration for a number of publications \cite{diaphony,leeb,jhk,hk1,hk2,hk3},
and have been for this paper. 

In order to assess the quality of a certain point set and to decide whether it 
will do better than a typical `random-point' set as in Monte Carlo, its 
discrepancy has to be computed and has to be compared with that of sets of 
random points. 
This means that one has to know the probability distribution of the 
discrepancy under sets of random points. 
In \cite{hk1,hk2,hk3} the problem of calculating this distribution has 
been tackled for large classes of discrepancies. One of the techniques used 
in these papers was that of Feynman diagrams as mnemonics in the organization 
of the combinatorics in the calculations. This is a method originally 
introduced for calculations in quantum field theory (QFT), and besides this
technique, other resemblances with QFT were noticed in the mentioned papers.

In this paper we show how the problem of calculating the probability 
distribution of a quadratic discrepancy, defined as an average-case complexity, 
can be cast in the form of a problem in terms of a QFT from the start. 
In particular, 
we show how the distribution can be calculated as a perturbation series in 
$1/N$ on the 
distribution for asymptotically large number of points $N$ in the point set. 
As examples, we use the Lego problem class and the Wiener problem class 
(defined in, for example, \cite{hkh1} and in this paper) to 
apply this method.

\section{General formalism}
We shall always take the integration region to
be the $s$-dimensional unit hypercube $\Kube=[0,1)^s$. The point set $X_N$
consists of $N$ points $x_k^\nu$, where $k=1,2,\ldots,N$ labels the
points and $\nu=1,2,\ldots,s$ their co-ordinates.
Defined as an average-case complexity on a class of functions 
$\phi:\Kube\mapsto\Real$ with measure $\mu$, the discrepancy $D_N$ of the point
set $X_N$ is given by 
\begin{equation}
   D_N \;=\; N\int\eta^2_N[\phi]\,\Dp \quad,\quad
   \eta_N[\phi] \;=\; \frac{1}{N}\sumN \phi(x_k) - \int_\Kube \phi(x)\,dx \;\;.
\label{induceddiscr}   
\end{equation}
When $X_N$ consists of uniformly distributed random points, 
then the discrepancy $D_N$ is a 
random variable with a certain probability distribution $H$. 
This probability distribution has been calculated for different discrepancies 
in various publications \cite{jhk,hk1,hk2}, in which the generating 
function 
\begin{equation}
   G(z) \;=\; \Exp{e^{zD_N}}
\end{equation}
has been used, where $\mathsf{E}$ denotes the expectation value of a random 
variable.   This paper will also concentrate on the calculation 
of $G(z)$. Given $G$, the probability density $H$ can then be calculated 
by the Laplace transform 
\begin{equation}
   H(D_N=t) \;=\; \frac{1}{2\pi i}\intinfi e^{-zt}G(z)\,dz \;\;.
\label{Laplace}   
\end{equation}
From now on, we assume the measure $\mu$ to be Gaussian, and propose
the calculation of the generating function from an explicit expression 
in terms of $\mu$, which we will now derive.
The integration error $\eta_N[\phi]$ can be written as a contraction 
$\eta_N[\phi]=\int \de\eta_N(x)\phi(x)\,dx$ of the function $\phi$ with a 
distribution given by
\begin{equation}
   \de\eta_N(x) \;=\; \frac{1}{N}\sumN\left[\de(x-x_k) - 1\right] \;\;,
\label{defJ}   
\end{equation}
where $\de(x-x_k)$ represents the $s$-dimensional Dirac $\de$-distribution 
in $\Kube$. In terms of the distribution $\de\eta_N$, the discrepancy is given 
by 
\begin{equation}
   D_N \;=\; N\intkk \de\eta_N(x)\twop(x,y)\de\eta_N(y)\,dxdy \;\;,
\end{equation}
where $\twop$ is the two-point Green function of the measure $\mu$:
\begin{equation}
   \twop(x,y)
   \;=\; \int \phi(x)\phi(y)\,\Dp \;\;.
\label{twopoint}   
\end{equation}
Notice that $G(0)$ has to be equal to one in order for the probability 
distribution $H$ to be normalized to one. 
This means that $\int\Dp=1$ and that \eqn{twopoint} 
indeed is the proper definition of the two-point Green function. 
Because we assume the measure to be Gaussian, 
we can write for the generating function
\begin{align}
   G(z) 
   &\;=\; \int_{\Kube^N}\exp(zD_N)\,dx_1\cdots dx_N \\
   &\;=\; \int_{\Kube^N}
          \int\exp\left(\sqrt{2zN}\int \de\eta_N(y)\phi(y)\,dy\right)\Dp\;
	  dx_1\cdots dx_N \;\;.
\end{align}
If now the definition (\ref{defJ}) of $\de\eta_N(y)$ is substituted, and the 
integrals over $x_1,\ldots,x_N$ are performed, we arrive at 
\begin{equation}
   G(z) 
   \;=\; \int \left(\intk e^{g[\phi(x)-\intk \phi(y)\,dy]}\,dx\right)^N \Dp 
         \quad,\quad  g \;=\; \sqrt{\frac{2z}{N}} \;\;.
\label{genG}
\end{equation}
For the Lego problem class and the Wiener problem class alternative derivations 
are given in Appendix A.

\subsection{The path integral}
\subsubsection{The action}
Using \eqn{genG}, the generating function can be written as a Euclidean 
path integral (cf. \cite{Rivers}) with an action $S$ given by 
\begin{equation}
   S[\phi] 
   \;=\; \frac{1}{2}\intkk \phi(x)\vLa(x,y)\phi(y)\,dxdy 
         - N\log\left(\intk e^{g[\phi(x)-\intk \phi(y)\,dy]}\,dx\right) \;\;,
\label{genaction}		    
\end{equation}
where $\vLa$ is the symmetric linear operator with boundary conditions 
which is the inverse of the two-point Green function under the measure $\mu$:
\begin{equation}
   \intk \vLa(x_1,y)\twop(y,x_2)\,dy
   \;=\; \de(x_1-x_2) \;\;.\label{defDelta} 
\end{equation} 
The two-point Green function $\twop$ satisfies the boundary conditions with
both of its arguments. From now on, we will 
assume the boundary conditions to be included in $\vLa$. Formally this can 
be realized by adding linear operators with $\de$-distributions centered 
around the boundaries, multiplied with an arbitrary large number~\footnote{For 
example $\vLa(x,y)=\lim\limits_{M\ra\infty}\left[-\frac{d^2}{dx^2}
+M\de(x-\frac{1}{3})\frac{d}{dx} +M\de(x-\frac{1}{2})\right]\de(x-y)$ 
is the linear operator $-\frac{d^2}{dx^2}$ with boundary conditions 
$\frac{d\phi}{dx}(\frac{1}{3})=\phi(\frac{1}{2})=0$.}. The large numbers 
guarantee that functions which do not satisfy the boundary conditions give no 
contribution to the path integral. 
Notice that, because Gaussian measures are completely defined by their 
two-point Green function, $\vLa$ can be used as a definition of $\mu$; 
functional integrals under $\mu$ can be written as path integrals with 
an action given by 
\begin{equation}
   S_0[\phi]
   \;=\; \frac{1}{2}\intkk \phi(x)\vLa(x,y)\phi(y)\,dxdy \;\;. 
\end{equation}

\subsubsection{Gaussian measures on a countable basis}
In \cite{hkh1} it has been pointed out that a large class of quadratic 
discrepancies, including the $L_2^*$-discrepancy in any dimension, can be 
constructed with a Gaussian measure on a class of functions defined by a 
countable set of basis functions. In this paper, we will further only consider 
function classes of this kind. We assume that the members $\phi$ of the class 
can be written as linear combinations 
\begin{equation}
   \phi(x) 
   \;=\; \sumn \phi_nu_n(x) \;\;,\quad \phi_n\in\Real
\end{equation}
of a countable set of basis functions $\{u_n\}$. 
Products of the basis functions are assumed to be integrable. In particular, 
we assume that the parameters
\begin{equation}
  w_n 
  \;=\; 
  \intk u_n(x)\,dx 
  \quad\textrm{and}\quad 
  a_{m,n} 
  \;=\; 
  \intk u_m(x)u_n(x)\,dx
\end{equation}
exist. 
On such a class of functions a Gaussian measure is defined by taking 
\begin{equation}
  \Dp 
  \;=\;
  \prod_{n}\frac{\exp(-\phi_n^2/2\si_n^2)}{\sqrt{2\pi\si_n^2}}\,d\phi_n
  \;\;,\quad \si_n\in\Real \;\;.
\label{accdefdmu}  
\end{equation}
For the measure to be suitably defined, the strengths $\si_n$ have to
satisfy certain restrictions which can be translated into the 
requirement that $\Exp{D_N}$ exists. 
With this measure, the discrepancy becomes
\begin{equation}
   D_N
   \;=\; \frac{1}{N}\dsumN\distwop(x_k,x_l) \;\;,\quad 
   \distwop(x_k,x_l)
   \;=\; \sum_n\si_n^2(u_n(x_k)-w_n)(u_n(x_l)-w_n) \;\;. 
\end{equation}
A connection with the foregoing can be established with the remark that 
in this case, $\twop$ as well as $\vLa$ can be written in terms of the basis: 
\begin{equation}
   \twop(x_1,x_2)
   \;=\; \sum_n\si_n^2u_n(x_1)u_n(x_2) \;\;,\quad
   \vLa(x_1,x_2)
   \;=\; \sum_n\frac{1}{\si_n^2}\,u_n(x_1)u_n(x_2) \;\;. 
\label{speccandDe}   
\end{equation}
The basis consists of the eigenfunctions of $\vLa$ and the strengths correspond 
with the eigenvalues:
\begin{equation}
   \intk\vLa(x,y)u_n(y)\,dy
   \;=\; \la_nu_n(x) \;\;,\quad \frac{1}{\si_n^2}=\la_n \;\;.
\end{equation}
The boundary conditions are those satisfied by the basis functions. 
Notice that the restriction to such classes of functions is 
equivalent with the restriction to measures defined with operators 
$\vLa$ that allow for a spectral decomposition in terms 
of their eigenfunctions as in \eqn{speccandDe}. 
Following the notation that is more frequently used in the path integral 
formulation of QFT, the measure can be written as 
\begin{equation}
   d\mu[\phi] 
   \;=\; \exp(-S_0[\phi])\,\Dpath\phi \;\;,
\end{equation}
but whenever accurate analyses are needed, we will refer to \eqn{accdefdmu}.

\subsection{Perturbation theory}
The action given by \eqn{genaction} is highly non-local because it multiplies 
function values $\phi(x)$ and $\phi(y)$ at finite distances $|x-y|$. 
In this respect the similarity with ordinary QFT fails. However, 
because we are mainly interested in $H(t)$ as an asymptotic 
expansion in $1/N$ and we assume $N$ to be very large, 
it can be written as a 
perturbation series in $1/N$ and locality is restored if the series is 
truncated at finite order. The zeroth-order term of the action will then, as in
QFT, be quadratic, and the remainder, denoted by $V$, will
have an expansion starting with terms of $\Ord(\phi^3)$:
\begin{equation}
   S[\phi] \;=\; \frac{1}{2}\intkk \phi(x)A_z(x,y)\phi(y)\,dxdy + V[\phi]\;\;,
\end{equation}  
\inttext{1.0}{0.5}{with}
\begin{equation}
   A_z(x,y) \;=\; \vLa(x,y) - 2z\de(x-y) + 2z \label{defA} \;\; 
\end{equation}  
\inttext{1.3}{0.5}{and}
\begin{align}
   V[\phi] &\;=\; NgJ_1 + N\half g^2\left(J_2 - J_1^2\right)
	  	  - N\log\left(\intk e^{g\phi}\,dx\right) \\
	   &\;=\; Ng^3\left( -\sfrac{1}{6}J_3
	                     +\sfrac{1}{2}J_1J_2
			     -\sfrac{1}{3}J_1^3 \right) 
			     + N\Ord(g^4)
	   \;\;,\quad J_p=\intk\phi(x)^p\,dx \notag\;\;.		     
\end{align}
Perturbation theory can now be applied to calculate $G(z)$. The total path 
integral (\ref{genG}) is evaluated as a perturbation series in $1/N$ on the 
norm of a 
Gaussian measure $\mu_z$, naively defined by
\begin{equation}
   d\mu_z[\phi]
   \;=\; \exp\left(-\frac{1}{2}\intkk\phi(x)A_z(x,y)\phi(y)\,dxdy\right)\,
         \Dpath\phi 
\end{equation}
(notice that $A_z$ inherits the boundary conditions included in $\vLa$). 
The series can be written as a diagrammatic expansion with a {\em propagator}
$\prop_z$, which is the two-point function of the measure $\mu_z$, so 
\begin{equation}
   \prop_z(x_1,x_2)
   \;=\; \frac{\int\phi(x_1)\phi(x_2)\,d\mu_z[\phi]}{\int d\mu_z[\phi]}
\end{equation}
and it satisfies
\begin{equation}
   \intk A_z(x_1,y)\prop_z(y,x_2)\,dy 
   \;=\; \de(x_1-x_2) \;\;.
\end{equation}
The diagrams are a help in the organization of the terms that contribute to a
given order in the perturbation series. These terms are proportional to moments 
$\int\phi(x_i)\cdots\phi(x_j)\,d\mu_z[\phi]$, integrated over the various 
co-ordinates $x_i,\ldots,x_j$, where the order of the moment is equal to the 
order of the term in the perturbation series.

According to the Gaussian integration rules, the zeroth order term in the 
series is proportional to $(\det A_z)^{-1/2}$. It is, however, not clear at
this point what the remaining factor is and whether $\det A_z$ is defined
properly. To overcome this problem, 
we will assume that the operator $A_z$ has a spectral 
decomposition in terms of its eigenfunctions. Because
$A_z\ra\vLa$ if $z\ra0$, we know that for every eigenvalue $\la_n$ of $\vLa$
there is 
an eigenvalue $\la_n(z)$ of $A_z$ with $\la_n(0)=\lim_{z\ra0}\la_n(z)=\la_n$.
As a result of this and the definition of $\mu_z$, the zeroth order term 
$G_0(z)=\int d\mu_z[\phi]$ is given by 
\begin{equation}
   G_0(z)
   \;=\; \prod_n\left(\sqrt{\frac{\la_n(0)}{2\pi}}\int
                    \exp(-\half\la_n(z)\phi_n^2)\,d\phi_n \right)
   \;=\; \left(\prod_n\frac{\la_n(0)}{\la_n(z)}\right)^{1/2} .
\label{G0}
\end{equation}
If we denote the eigenfunction corresponding with the eigenvalue $\la_n(z)$ 
by $u_{n,z}$, then the propagator is given by 
\begin{equation}
   \prop_z(x_1,x_2)
   \;=\; \sum_n\frac{1}{\la_n(z)}\,u_{n,z}(x_1)u_{n,z}(x_2) \;\;.
\end{equation}
The higher orders orders in the perturbation series consist of convolutions 
of the propagator, multiplied with the zeroth order term. From now on, 
the index $z$ in $A_z$ and $\prop_z$ will be omitted.

\subsection{Gauge freedom}
An interesting feature of the propagator is that it is not unique. This is a 
result of the fact that a global translation 
\begin{equation}
   \vTh_c:\;\phi(x)
   \mapsto (\vTh_c\phi)(x)=\phi(x)+c \;,\;\; c\in\Real 
\end{equation}
only results in a change of the action that is at most 
linear in the functions $\phi$: 
\begin{equation}
   S[\vTh_c\phi]
   \;=\; S[\phi]+2\al c\chi[\phi]+\half\al c^2 \;\;,
\end{equation}
where $\chi$ is such that $\chi[\vTh_c\phi]=\chi[\phi]+c$ and where $\al>0$. 
In terms of the original integration problem this can be explained by the fact 
that the integration error is the same for two integrands that differ by a 
constant and the fact that $\mu$ is Gaussian.
As a result of this, and the fact that we may assume the 
`measure' $\Dpath\phi$ to be invariant under $\vTh_c$, we can take 
any functional $\xi$ linear in $\phi$ with $\xi[\vTh_c\phi]=\xi[\phi]+c$ and 
write 
\begin{align}
   \int \exp(-S[\phi])\,\Dpath\phi
  &\;=\; \int \de(\xi[\phi]-c)\,\exp(-S[\phi])\,dc\,\Dpath\phi \\
  &\;=\; \sqrt{\frac{2\pi}{\al}}\int \de(\xi[\phi])\,\exp(-S_\vTh[\phi])\,
         \Dpath\phi \;\;,
\end{align}	 
\inttext{1.0}{0.5}{with}
\begin{equation}
   S_\vTh[\phi]
   \;=\;S[\phi]-\half\al\chi^2[\phi] \;\;.
\end{equation}
The $\de$-distribution in the last expression tells us that the variable 
$\xi[\phi]$ `decouples' from its perpendicular directions in $\phi$-space. 
This decoupling of one degree of freedom is necessary in order for the total 
integral to exist, because the new action 
$S_\vTh$ is invariant under global translations. 
Now we add to this action a function $F$, such that 
\begin{equation}
   I[F]
   \;=\; \int_{-\infty}^{\infty}\exp(-F(c))\,dc 
\label{defI}   
\end{equation}
exists, so that 
\begin{equation}
   \int \exp(-S[\phi])\,\Dpath\phi
   \;=\; \frac{1}{I[F]}\sqrt{\frac{2\pi}{\al}}\int 
         \exp(-F(\xi[\phi]) - S_\vTh[\phi])\, \Dpath\phi \;\;.
\end{equation}
The result is that the total integral expressed in terms of $S[\phi]$ is, 
apart from some normalizations, completely equivalent with the integral 
expressed in terms of $F(\xi[\phi])-\half\al\chi^2[\phi]+S[\phi]$. 
Because of 
this freedom in the choice of the action there is a 
freedom in the choice of $A$ and the propagator $\prop$, which 
we call the {\em gauge freedom}. 
As usual, under different gauges, individual Feynman diagrams evaluate to
different results, but the perturbation series as a whole is gauge-invariant.

Notice that, in the orthogonal-basis picture, a change of gauge in general 
results in a change of the basis functions and the eigenvalues. However, 
the zeroth order term in the perturbation series, for example defined as in 
\eqn{G0}, has to be gauge invariant.

\subsection{Instantons\label{geninst}}
An expansion of the action to evaluate the generating function only makes sense
when it is an expansion around a minimum, so that it represents a saddle point 
approximation of the path integral. Therefore, a straightforward expansion 
such as just proposed, which is in fact an expansion around the trivial 
solution $\phi=0$, 
is only correct if it is an expansion around the minimum of 
the action, that is, if the trivial solution gives the only minimum of the 
action. General 
extrema of the action are given by solutions of the {\em field equation}
\begin{equation}
   \intk\vLa(x,y)\phi(y)\,dy + Ng - Ng\frac{e^{g\phi(x)}}{\intk e^{g\phi(y)}\,dy}
   \;=\; 0 \;\;.
\label{fieldeqn}    
\end{equation}
Depending on the value of $z$, non-trivial solutions may also exist. 
At this point it can be said that, because 
$\vLa(x,y)$ as well as $\phi(x)$ is real, non-trivial
solutions only exist if $z$ is real and non-zero so that $g\in\Real$. 
In the analysis 
of the solutions we therefore can do a scaling $\phi(x)\mapsto \phi(x)/g$
so that the action for these solutions is given by
\begin{equation}
   \Si[\phi]\;\equiv\;
   \frac{S[\frac{1}{g}\phi]}{N} 
   \;=\; \frac{1}{2}\,\frac{\intk \phi(x)e^{\phi(x)}\,dx}{\intk e^{\phi(y)}\,dy}
         + \frac{1}{2}\intk \phi(x)\,dx
         - \log\left(\intk e^{\phi(x)}\,dx\right)\;\;.
\label{instact}	 
\end{equation}
These non-trivial solutions we call instantons (cf. \cite{Coleman}), 
although this may not be a rigorously correct nomenclature, in the field 
theoretical sense, for all situations we will encounter. 
Notice that instantons under different gauges only differ by a constant. This 
is easy to see because instantons under a general gauge, characterized by 
$F$ and $\xi$, are equal to solutions $\phi_\vTh$ of the 
field equation obtained from the gauge invariant action $S_\vTh$ plus a constant 
$c$ determined by the relation $F^\pr(\xi[\phi_\vTh]+c)=0$. The values of $z$ 
for which they appear and the value of the action are gauge invariant, 
as can be concluded from \eqn{fieldeqn} and \eqn{instact}.

If $N$ becomes large, then the contribution of an instanton to the path 
integral will behave as $e^{-N\Si[\phi]}$, where $\Si[\phi]$ does not 
depend on $N$ (Notice that $\phi(x)$ does not depend on $N$ because the field 
equation for these rescaled functions does not depend on $N$.). 
The $e^{-N\Si[\phi]}$-like behavior of the instanton contribution makes it 
invisible in the perturbative expansion around $1/N=0$. If $\Si[\phi]$ is 
larger than zero, this will not be a problem, because the contribution will be 
very small. If, however, $\Si[\phi]$ is equal to zero, then the contribution 
will be more substantial, and it will even explode if $\Si[\phi]$ is 
negative~\footnote{Notice that, to be able to do make a perturbation series 
around $\phi=0$, the action has to be zero for this solution, for else the 
terms would all become zero or would explode for large $N$.}.
This would really be a major 
problem, if it were not for the fact that, in the cases we encounter, $z$ 
has to be real and larger than zero for these instantons to exist, and,  
according to \eqn{Laplace}, we want to integrate $G(z)$ along the imaginary 
$z$-axis. In the end, when we want to close the 
integration contour in the complex $z$-plane to the right, we might meet the 
problem again. However, the function we want to integrate is an expansion in 
$1/N$ of the generating function, which is also an expansion in $z$ around 
$z=0$ that can be integrated term by term, and therefore we will never face the
infinite instanton contributions.

\section{The Lego problem class}
\subsection{Definition}
The Lego problem class is obtained by dissecting the hypercube $\Kube$ into $M$ 
non-overlap\-ping bins and by taking the characteristic functions $\vt_n$ of 
the bins as the basis functions. Then $w_n$ is the volume of bin $n$ and this 
implies that all $w_n$ are larger than zero. The functions $\vt_n$ and the 
weights $w_n$ moreover satisfy 
\begin{equation}
   \vt_n(x)\vt_m(x) = \de_{n,m}\vt_n(x) \;\;,\quad
   \sumM \vt_n(x) = 1\;\;\forall~x\in\Kube \quad\textrm{and}\quad
   \sumM w_n = 1 \;\;.
\label{relchar}   
\end{equation}
The coefficients $a_{n,m}$ are equal to $w_n\de_{n,m}$.
Notice that, for this function class, the number of basis functions is not only 
countable but even finite. 

In the following, we restrict the strengths $\si_n$ such that 
all $\si_m^2w_m$ are equal to $1$. This choice models functions in which 
the largest fluctuations appear over the smallest intervals. 
Although not a priori attractive in many cases, this choice is 
quite appropriate for particle physics, where cross sections 
display precisely this kind of behavior. Moreover, the average-case 
complexity is the same as that of the $\chi^2$-goodness-of-fit test. 
With this choice, the discrepancy becomes
\begin{equation}
   D_N
   \;=\; \frac{1}{N}\dsumN\sumM\frac{\vt_n(x_k)\vt_n(x_l)}{w_n} - N \;\;.
\label{legodiscr}   
\end{equation}
In \cite{hkh1} it has been shown that, for asymptotically large $N$, the 
probability distribution of $D_N$ under truly random point sets approaches a 
Gaussian distribution whenever $M\ra\infty$. 

\subsection{The action}
Using \eqn{genG} and the relations (\ref{relchar}), we can 
write down the generating function without facing any ambiguities, obtaining
\begin{equation}
  G(z) 
   \;=\;  \prodM\left(\frac{w_n}{2\pi}\right)^{\frac{1}{2}}
          \int \exp\left(-S[\phi]\right)\,d^M\!\phi \;\;,
          \label{Legogen} 
\end{equation}
\inttext{1.0}{0.5}{with}	  
\begin{equation}
   S[\phi]
   \;=\;  \frac{1}{2}\sumM w_n\phi_n^2
          + Ng\sumM w_n\phi_n
          - N\log\left(\sumM w_ne^{g\phi_n}\right) \;\;,\;\;
   g = \sqrt{\frac{2z}{N}} \;\;,
\label{legoaction}      
\end{equation}
where $\Real^M\ni\phi=(\phi_1,\ldots,\phi_M)$ and the integration region 
extends over the whole of $\Real^M$. In section~\ref{Legoinstantons} it will 
be shown that,
if $\textrm{Re}~z<\half w_{\rmin}$ with 
$w_{\rmin}=\min_nw_n$, then the only extremal point 
of the action is a minimum at $\phi_n=0$, $n=1,\ldots,M$, so that the saddle 
point approximation boils down to a straightforward expansion in $1/N$. 
The action can be 
written in terms of a symmetric linear operator $A$ and a potential $V$ 
as~\footnote{Summations without without explicit limits from $1$ to $M$.}
\begin{equation}
   S[\phi]
   \;=\; \half(\phi,A\phi)+V[\phi] \;\;,
\end{equation}
\inttext{1.3}{0.5}{with}
\begin{align}
  &A_{n,m} 
   \;=\; (1-2z)w_n\de_{n,m} + 2zw_nw_m \;\;,\\
  &V[\phi]
   \;=\;   Ng\sum w_n\phi_n 
         + z\sum w_n\phi_n^2 
         - z\left(\sum w_n\phi_n\right)^2 
         - N\log\left(\sum w_ne^{g\phi_n}\right) \;\;,
   \label{Legopot}
\end{align}
where $(\cdot,\cdot)$ stands for the canonical inproduct. 
In order for $A$ to define the Gaussian measure well, 
its real part has to have $M$ positive eigenvalues~\footnote{$A$ does 
not have to be a mapping $\Real^M\mapsto\Real^M$; because $z$ may be complex 
it is a linear transformation $\Real^M\mapsto\Comp^M$.}. We assumed that 
$\textrm{Re}~z\in(0,\half w_{\rmin}]$ with $w_{\rmin}\leq\half$, 
so that Appendix B makes clear that the real part of $A$ indeed has $M$ 
positive 
eigenvalues. 

To calculate the zeroth 
order term in an $1/N$ expansion of the path integral, 
the determinant of $A$ has to be calculated. The result is that 
\begin{equation}
   G_0(z)
   \;=\;  \prodM\left(\frac{w_n}{2\pi}\right)^{\frac{1}{2}}
          \sqrt{\frac{(2\pi)^M}{\det A}}
   \;=\;  \frac{1}{(1-2z)^{\frac{M-1}{2}}} \;\;,   
\end{equation}
which is precisely the moment generating function for the $\chi^2$-distribution 
with $M-1$ degrees of freedom.
Now perturbation theory can be applied to calculate the rest of the expansion. 
Therefore Feynman diagrams can be used with a propagator given by the inverse 
matrix $A^{-1}$ of $A$:
\begin{equation}
   A^{-1}_{n,m}
   \;=\; \frac{1}{1-2z}\left[\frac{\de_{n,m}}{w_n}-2z\right] \;\;.
\end{equation}

\subsection{Gauge freedom}
Because of the gauge freedom, 
the path integral can equally well be defined with an 
action given by $F(\xi[\phi])-\half\xi^2[\phi]+S[\phi]$, 
where $\xi[\phi]=\sum w_n\phi_n$ and where $F$ can be any 
function with the restriction that $I[F]$ (\eqn{defI}) exists. 
We can translate this into a freedom in 
the choice of the propagator $A^{-1}$, for if we take $F(\xi)$ as a series in 
$\xi$, then the quadratic part will change the propagator. If we, for example, 
take $F(\xi)=\half[1+2z(\frac{1}{\vp}-1)]\xi^2$ with $\vp>0$, then  
\begin{align}
  &A_{n,m} \;\mapsto\; A_{n,m}(\vp)
   \;=\; (1-2z)w_n\de_{n,m} + \frac{2z}{\vp}\,w_nw_m \;\;,\\
  &A_{n,m}^{-1} \;\mapsto\; A_{n,m}^{-1}(\vp) 
   \;=\; \frac{1}{1-2z}\left[\frac{\de_{n,m}}{w_n}
                                   - \frac{2z}{\vp(1-2z)+2z}\right] \;\;.
\end{align}
There are two interesting limits for $\vp$. The limit of $\vp\ra\infty$ 
(the `Feynman gauge') results in a diagonal propagator, which makes Feynman 
calculus easy. The limit of $\vp\ra 0$ (the `Landau gauge') results 
in a singular propagator, that is, $A(0)$ is not defined.

\subsection{Instantons\label{Legoinstantons}}
We start this section with a repetition of the statement that non-trivial
instanton solutions only exist if $z\in[0,\infty)$ (section \ref{geninst}).
In order to investigate the instantons in the Lego problem class, we analyze 
the 
action in terms of the variables $y_n=g\phi_n+2z$, that is, we consider the 
integral $\int_{\Real^M}\exp(-N\Si[y])\,d^M\!y$\,, with 
\begin{equation}
   \Si[y]
   \;=\; z + \frac{1}{4z}\sum w_ny_n^2 - \log\left(\sum w_ne^{y_n}\right) \;\;.
\end{equation}
We are interested in the minima of $\Si$. The `perturbative' minimum 
$\phi_n=0$, $n=1,\ldots,M$ corresponds to $y_n=2z$, $n=1,\ldots,M$, and general
extrema of $\Si$ are situated at points 
$y$ which are solutions of the equations
\begin{equation}
   \frac{\partial\Si}{\partial y_k}(y)
   \;=\; 0 
   \quad\Leftrightarrow\quad
   \frac{e^{y_k}}{y_k}
   \;=\; \frac{1}{2z}\sum w_ne^{y_n} \;\;,\quad k=1,\ldots,M \;\;.
\label{extremal}   
\end{equation}
If $z$ is positive, $e^{y_k}/y_k$, and therefore 
$y_k$, has to be positive for every $k$. 
The result is that the 
$y_k$ can take at most two values in one solution $y$ (\fig{yminyplus}). 
If they all 
take the same value, this value is $2z$, and we get the perturbative solution.
If they take two values, one of them, $y_+$, is larger that $1$ and the
other, $y_-$, is smaller than $1$. 
With these results, and the fact that \eqn{extremal} implies that 
\begin{equation}
   \sum w_ny_n 
   \;=\; 2z \;\;,
\label{2zeqn}   
\end{equation}
we see that there are no solutions but the perturbative one if 
$2z<w_{\rmin}$, where $w_{\rmin}=\min_nw_n$. 
\begin{figure}[t]
\begin{center}
\begin{picture}(150,150)(0,0)
\LinAxis(0,0)(150,0)(1,1,3,0,1.5)
\LinAxis(0,0)(0,150)(1,1,-3,0,1.5)
\LinAxis(0,150)(150,150)(1,1,-3,0,1.5)
\LinAxis(150,0)(150,150)(1,1,3,0,1.5)
\Line(50,0)(50,150)
\Line(0,60)(150,60) 
\DashLine(50,90)(150,90){1.5}\Text(40,90)[]{$e^v$}
\DashLine(59,60)(59,90){1.5}\Text(60,52)[]{$y_-$}
\DashLine(103.8,60)(103.8,90){1.5}\Text(104.8,52)[]{$y_+$}
\Text(  0.0,-10)[]{$-2$}
\Text( 50.0,-10)[]{$0$}
\Text(150.0,-10)[]{$4$}
\Text(-12,  0.0)[]{$-8$}
\Text(-7, 60.0)[]{$0$}
\Text(-10,150)[]{$12$}
\Text(75,-20)[]{$t$}
\Text(-35,75)[]{$f(t)$}
\Text(80,25)[l]{$f(t)=\displaystyle\frac{e^t}{t}$}
\Curve{(  0.0, 59.5) (  1.5, 59.4) (  3.0, 59.4) (  4.5, 59.3) (  6.0, 59.3)
       (  7.5, 59.2) (  9.0, 59.1) ( 10.5, 59.0) ( 12.0, 58.9) ( 13.5, 58.8)
       ( 15.0, 58.7) ( 16.5, 58.5) ( 18.0, 58.4) ( 19.5, 58.2) ( 21.0, 58.0)
       ( 22.5, 57.7) ( 24.0, 57.5) ( 25.5, 57.1) ( 27.0, 56.8) ( 28.5, 56.3)
       ( 30.0, 55.8) ( 31.5, 55.2) ( 33.0, 54.4) ( 34.5, 53.5) ( 36.0, 52.3)
       ( 37.5, 50.9) ( 39.0, 49.0) ( 40.5, 46.5) ( 42.0, 43.0) ( 43.5, 37.8)
       ( 45.0, 29.3) ( 46.5, 13.4) (47,0)}
\Curve{( 52,150)
       ( 53.0,130.5) ( 54.5,109.9) ( 56.0, 99.7) ( 57.5, 93.7) ( 59.0, 89.9)
       ( 60.5, 87.2) ( 62.0, 85.3) ( 63.5, 83.8) ( 65.0, 82.8) ( 66.5, 82.0)
       ( 68.0, 81.4) ( 69.5, 81.0) ( 71.0, 80.7) ( 72.5, 80.5) ( 74.0, 80.4)
       ( 75.5, 80.4) ( 77.0, 80.4) ( 78.5, 80.6) ( 80.0, 80.8) ( 81.5, 81.0)
       ( 83.0, 81.3) ( 84.5, 81.6) ( 86.0, 82.0) ( 87.5, 82.4) ( 89.0, 82.9)
       ( 90.5, 83.4) ( 92.0, 84.0) ( 93.5, 84.6) ( 95.0, 85.2) ( 96.5, 85.9)
       ( 98.0, 86.6) ( 99.5, 87.4) (101.0, 88.3) (102.5, 89.2) (104.0, 90.1)
       (105.5, 91.1) (107.0, 92.2) (108.5, 93.3) (110.0, 94.4) (111.5, 95.7)
       (113.0, 97.0) (114.5, 98.4) (116.0, 99.8) (117.5,101.3) (119.0,102.9)
       (120.5,104.6) (122.0,106.4) (123.5,108.3) (125.0,110.2) (126.5,112.3)
       (128.0,114.4) (129.5,116.7) (131.0,119.1) (132.5,121.6) (134.0,124.3)
       (135.5,127.0) (137.0,130.0) (138.5,133.0) (140.0,136.2) (141.5,139.6)
       (143.0,143.2) (144.5,146.9) (145.6,150)}
\end{picture}    
\vspace{25pt}
\caption{$y_-$ and $y_+$.}
\label{yminyplus}
\end{center}
\end{figure}

In the next section, the other extremal points will be analyzed and it will 
appear that minima occur with $\Si[y]<0$. This means 
that, in the limit of $N\ra\infty$, the integral of $\exp(-N\Si)$ is not 
defined; there is a `wall' in the complex $z$ plane along the positive real 
side of the imaginary axis, to the right of which the generating function is
not defined. That this is not an artifact of our approach, can be 
seen in the expression of the generating function given by \eqn{multinom} in 
Appendix A. It is shown there that the generating function is not defined 
if $\textrm{Re}~z>\frac{w_n}{w_n-1}\log w_n$ for any one of the $w_n$.

We know that, on the perturbative level, the generating function has a 
singularity at $z=\half$, but the instanton contributions cannot correspond 
with it, because they will appear already for $\textrm{Re}~z<\half$. 
However, in order to calculate the probability density $H$ with the Laplace 
transform, using the perturbative expression of $G(z)$, we can just calculate 
the contribution of the singularity at $z=\half$, for {\em that is} the 
contribution to the perturbative expansion of $H(t)$.

\subsubsection{The Wall}
To expose the nature of the extrema of $\Si$, we have to investigate the 
eigenvalues $\la$ of the second derivative matrix $\vDe$ of $\Si$ in the 
extremal points. This matrix is given by 
\begin{equation}
   \vDe_{k,l}(y)
   \;\equiv\; \frac{\partial^2\Si}{\partial y_k\partial y_l}(y) 
   \;=\; a_k(y)\de_{k,l}+b_k(y)b_l(y)\;\;, 
\end{equation}
with
\begin{equation}
   a_k(y)=\frac{w_k}{2z}(1-y_k)\quad\textrm{and}\quad
   b_k(y)=\frac{w_ky_k}{2z}\;\;.
\end{equation}
To show that $\Si$ becomes negative, we only use its minima, 
and these correspond with 
extremal points in which all eigenvalues of $\vDe$ are positive. According to 
Appendix B, 
we are therefore only interested in cases where the degeneracy of 
negative $a_k$ is one, for else $\la=a_k$ would be a solution. We further are 
only interested in cases where there is only one negative $a_k$, for if there 
where more, say $a_k$ and $a_{k+1}$ with $a_k<a_{k+1}$, then there would be 
a solution $a_k<\la<a_{k+1}<0$. So we see that the only extremal points we 
are interested in have all co-ordinates $y_k$ equal, or have one 
$y_k=y_+$ and the others equal to $y_-$. If they are all
equal, then they have to be equal to $2z$, and for the extremal point to be a 
minimum $2z$ has to be smaller than $1$. This is the perturbative minimum. 
Whether the other extremal points are minima depends on whether 
$\det[\vDe]$ is positive in these points. The determinant can be written as 
\begin{equation}
   \det[\vDe(t)]
   \;=\; \frac{\prod_n w_n}{(2z)^{M+1}}\,(1-y_-)^{M-1}(y_+-1)
         \left(\frac{w_+y_+}{y_+-1} + \frac{(1-w_+)y_-}{y_--1}\right) \;\;.
\end{equation}
Now we notice that all extremal points can be labeled with a parameter $v$ by 
defining
\begin{equation}
   \frac{e^{y_\pm(v)}}{y_\pm(v)}
   \;=\; e^v \quad\textrm{with}\quad v\in(1,\infty) \;\;.
\end{equation}
We see that $y_\pm$ is a continuous and differentiable function of $v$ and we 
have that $dy_\pm/dv=y_\pm/(y_\pm-1)$. This parameterization induces 
a parameterization of $2z$, and with the help of \eqn{2zeqn} we see that
\begin{equation}
   \frac{d(2z)}{dv}
   \;=\; \frac{w_+y_+}{y_+-1} + \frac{(1-w_+)y_-}{y_--1} \;\;. 
\end{equation}
So we see that the sign of 
$\det[\vDe]$ is the same as the sign of $d(2z)/dv$: if an extremal point is a 
minimum, then $d(2z)/dv>0$. 
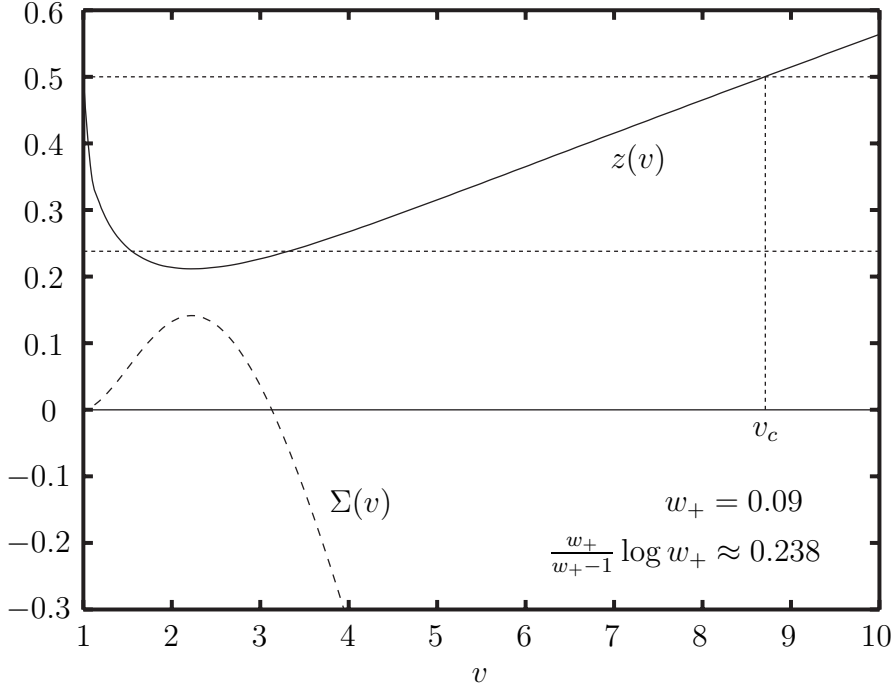
\begin{figure}[t]
\begin{center}
\begin{picture}(300,226)(0,0)
\LinAxis(0,0)(300,0)(9,1,3,0,1.5)
\LinAxis(0,0)(0,226)(9,1,-3,0,1.5)
\LinAxis(0,226)(300,226)(9,1,-3,0,1.5)
\LinAxis(300,0)(300,226)(9,1,3,0,1.5)
\Text(0,-10)[]{$1$}
\Text(33.3,-10)[]{$2$}
\Text(66.7,-10)[]{$3$}
\Text(100,-10)[]{$4$}
\Text(133.3,-10)[]{$5$}
\Text(166.7,-10)[]{$6$}
\Text(200,-10)[]{$7$}
\Text(233.3,-10)[]{$8$}
\Text(266.7,-10)[]{$9$}
\Text(300,-10)[]{$10$}
\Text(-17,0)[]{$-0.3$}
\Text(-17,25.1)[]{$-0.2$}
\Text(-17,50.2)[]{$-0.1$}
\Text(-13,75.3)[]{$0$}
\Text(-15,100.4)[]{$0.1$}
\Text(-15,125.6)[]{$0.2$}
\Text(-15,150.7)[]{$0.3$}
\Text(-15,175.8)[]{$0.4$}
\Text(-15,200.9)[]{$0.5$}
\Text(-15,226)[]{$0.6$}
\Text(210,170)[]{$z(v)$}
\Text(105,40)[]{$\Si(v)$}
\Text(220,40)[l]{$w_+=0.09$}
\Text(176,20)[l]{$\frac{w_+}{w_+-1}\log w_+\approx 0.238$}
\Text(150,-25)[]{$v$}
\Line(0,75.3)(300,75.3)
\DashLine(0,135.1)(300,135.1){1.5}
\DashLine(0,200.9)(300,200.9){1.5}
\DashLine(257,75.3)(257,200.9){1.5}\Text(258,67.3)[]{$v_c$}
\Curve{(  0.0,200.9) (  3.0,164.5) (  6.0,153.5) (  9.0,146.6)
       ( 12.0,141.7) ( 15.0,138.1) ( 18.0,135.3) ( 21.0,133.3)
       ( 24.0,131.7) ( 27.0,130.5) ( 30.0,129.6) ( 33.0,129.1)
       ( 36.0,128.7) ( 39.0,128.5) ( 42.0,128.5) ( 45.0,128.6)
       ( 48.0,128.9) ( 51.0,129.2) ( 54.0,129.6) ( 57.0,130.1)
       ( 60.0,130.7) ( 63.0,131.4) ( 66.0,132.1) ( 69.0,132.8)
       ( 72.0,133.6) ( 75.0,134.5) ( 78.0,135.3) ( 81.0,136.2)
       ( 84.0,137.1) ( 87.0,138.1) ( 90.0,139.1) ( 93.0,140.1)
       ( 96.0,141.1) ( 99.0,142.1) (102.0,143.1) (105.0,144.2)
       (108.0,145.2) (111.0,146.3) (114.0,147.4) (117.0,148.5)
       (120.0,149.6) (123.0,150.7) (126.0,151.8) (129.0,152.9)
       (132.0,154.0) (135.0,155.1) (138.0,156.2) (141.0,157.4)
       (144.0,158.5) (147.0,159.6) (150.0,160.7) (153.0,161.9)
       (156.0,163.0) (159.0,164.1) (162.0,165.3) (165.0,166.4)
       (168.0,167.5) (171.0,168.7) (174.0,169.8) (177.0,171.0)
       (180.0,172.1) (183.0,173.2) (186.0,174.4) (189.0,175.5)
       (192.0,176.6) (195.0,177.8) (198.0,178.9) (201.0,180.0)
       (204.0,181.1) (207.0,182.3) (210.0,183.4) (213.0,184.5)
       (216.0,185.7) (219.0,186.8) (222.0,187.9) (225.0,189.0)
       (228.0,190.2) (231.0,191.3) (234.0,192.4) (237.0,193.5)
       (240.0,194.7) (243.0,195.8) (246.0,196.9) (249.0,198.0)
       (252.0,199.1) (255.0,200.2) (258.0,201.4) (261.0,202.5)
       (264.0,203.6) (267.0,204.7) (270.0,205.8) (273.0,206.9)
       (276.0,208.0) (279.0,209.2) (282.0,210.3) (285.0,211.4)
       (288.0,212.5) (291.0,213.6) (294.0,214.7) (297.0,215.8)
       (300.0,216.9) }
\DashCurve{(  0.0, 75.3) (  3.0, 76.4) (  6.0, 78.6) (  9.0, 81.6)
       ( 12.0, 85.0) ( 15.0, 88.8) ( 18.0, 92.7) ( 21.0, 96.6)
       ( 24.0,100.2) ( 27.0,103.4) ( 30.0,106.2) ( 33.0,108.4)
       ( 36.0,109.9) ( 39.0,110.7) ( 42.0,110.8) ( 45.0,110.1)
       ( 48.0,108.7) ( 51.0,106.6) ( 54.0,103.7) ( 57.0,100.2)
       ( 60.0, 95.9) ( 63.0, 91.1) ( 66.0, 85.7) ( 69.0, 79.7)
       ( 72.0, 73.2) ( 75.0, 66.2) ( 78.0, 58.8) ( 81.0, 51.0)
       ( 84.0, 42.9) ( 87.0, 34.4) ( 90.0, 25.6) ( 93.0, 16.5)
       ( 96.0,  7.1) ( 98.2, 0)}{3} 
\end{picture}    
\vspace{25pt}
\caption{$\Si$ and $z$ for instanton solutions parameterized with $v$.}
\label{Sandzofv}
\end{center}
\end{figure}
The minimal value that 
$v$ can take to represent a solution is $1$, which corresponds to 
$y_+=y_-=1$ and $2z=1$. 
It is easy to see that $d(2z)/dv\ra-\infty$ if $v\da 1$ and $w_+<\half$, where  
$w_+$ is the value of the weight belonging to the co-ordinate with the value 
$y_+$.
This means 
that if $v$ starts from $v=1$ and increases, then it will represent solutions 
with $d(2z)/dv<0$, which are local maxima. We know that, if $v\ra\infty$, then
$y_-\ra0$, $y_+\ra\infty$ and $2z=w_+y_++(1-w_+)y_-\ra\infty$, so that 
$d(2z)/dv$ has to become larger than $0$ at some point. The first point where 
$2z$ becomes equal to $1$ again we call $v_c$, so $2z(v_c)=2z(1)=1$ 
(\fig{Sandzofv}).
Also the function $\Si$ itself can be written in terms of $z(v)$ in the 
extremal points. Therefore we use that
\begin{equation}
   \frac{d}{dv}[w_+y_+^2 + (1-w_+)y_-^2]
   \;=\; 4z + 4\frac{dz}{dv} \;\;
\end{equation}
and that $w_+y_+^2 + (1-w_+)y_-^2=1$ if $v=1$, so that 
\begin{equation}
   \Si(v)
   \;=\; z(v) + \frac{1}{z(v)}\int_1^v z(x)\,dx + 1 - \frac{1}{4z(v)} - v
         - \log[2z(v)] \;\;.
\end{equation}
Now the problem arises. From the previous analysis of $z(v)$ we know that, 
if $1\leq v\leq v_c$, then $z(v)<\half$ so that 
\begin{equation}
   \Si(v_c)
   \;=\; 1 - v_c + 2\int_1^{v_c}z(x)\,dx \;<\; 0 \;\;. 
\end{equation}
Furthermore, we find that 
\begin{equation}
  \frac{d\Si}{dv} 
  = \left[1-\frac{1}{4z^2}\left(w_+y_+^2+(1-w_+)y_-^2\right)\right]\frac{dz}{dv}
  = -\frac{w_+(1-w_+)(y_+-y_-)^2}{4z^2}\,\frac{dz}{dv} \;\;,
\end{equation}
so that also $d\Si/dv<0$ in $v_c$. So there clearly is a region in $[1,v_c]$ 
where $dz/dv>0$ and $\Si(v)<0$. This means that in the region 
$\half w_{\rmin} <z<\half$ 
there are instanton solutions with negative action. The situation is shown in 
\fig{Sandzofv} for $w_{\rmin}=0.09$. A region where $dz/dv>0$ and $S(v)<0$ is 
clearly visible in $[1,v_c]$.

\section{The $L_2^*$-discrepancy and the Wiener problem class}
\subsection{Definition}
The standard $L_2^*$-discrepancy \cite{nieder1} is defined as the squared 
integration error made by integrating the characteristic functions 
$\vt_y$ of hypercubes 
$[0,y^1)\times[0,y^2)\times\cdots\times[0,y^s)\subset\Kube$ 
with the point set $X_N$, averaged over $y$:
\begin{equation}
   \frac{D_N}{N} 
   \;=\; \intk \left(\frac{1}{N}\sumN\vt_y(x_k)-\prod_{\nu=1}^sy^\nu\right)^2dy
   \;\;, \quad \vt_y(x)=\prod_{\nu=1}^s\theta(y^\nu-x^\nu) \;\;.
\label{defL_2^*}   
\end{equation}
The Wo\'zniakowski lemma from Ref.~\cite{woz} states that it can be written as 
in \eqn{induceddiscr}, that is, as 
the squared integration error, averaged with respect to a variation of the 
Wiener sheet measure in which the functions are pinned down at 
$x=(1,1,\ldots,1)$ rather than at $x=(0,0.\ldots,0)$. The Wiener sheet measure 
itself is Gaussian with the two-point Green function given by
\begin{equation}
   \twop(x,y) \;=\; \prod_{\nu=1}^s\min(x^\nu,y^\nu) \;\;.
\end{equation}
In \cite{hkh1} it has been shown that it has a spectral representation in 
terms of a set of orthogonal functions on $\Kube$. In the case $s=1$, on which 
we shall concentrate here, these functions are the 
eigenfunctions of the linear operator~\footnote{Primes stand for derivatives.} 
$\phi\mapsto-\phi^{\pr\pr}$ in the space of functions 
$\phi:\Kube\mapsto\Real$ with boundary conditions $\phi(0)=\phi^\pr(1)=0$, 
and, in fact, the measure can be defined with an action given by 
$S[\phi]=-\half\intk \phi(x)\phi^{\pr\pr}(x)\,dx$, with $\phi(0)=\phi^\pr(1)=0$.
Usually the measure is written in terms of an action
\begin{equation}
   S_0[\phi]
   \;=\; \frac{1}{2}\intk\phi^\pr(x)^2 \,dx 
   \;\;,\;\;\textrm{with}\;\; \phi(0)=0 \;\;,
\label{bareWienaction}				   
\end{equation}
from which the other one can be obtained by partial integration.

\subsection{The action}
\eqn{genG} states that the generating function of the $L_2^*$-discrepancy in 
one dimension is given by a path integral with an action $S$ given by
\begin{equation}
   S[\phi]
   \;=\; \frac{1}{2}\intk\phi^\pr(x)^2\,dx 
         + Ng\intk\phi(x)\,dx - N\log\left(\intk e^{g\phi(x)}\,dx\right)
	 \;\;,\;\; g=\sqrt{\frac{2z}{N}}\;,
\label{WienG}	 
\end{equation}
where the functions $\phi$ satisfy the boundary condition $\phi(0)=0$. In 
section~\ref{Wieninstantons} it will be shown that if 
$\textrm{Re}~z<\half\pi^2$, then the only extremal point of the action is a 
minimum at $\phi=0$, so that a saddle point expansion boils down to an 
expansion in $1/N$.

\subsection{Gauge freedom}
The boundary condition can be included into \eqn{WienG} by adding 
a term $\half M\phi(0)^2$ with $M\ra\infty$. 
If $M$ is taken finite, then the problem class is not restricted anymore to 
functions with $\phi(0)=0$, but $\phi(0)$ gets a Gaussian distribution. For 
the discrepancy, however, this does not matter, because the extra functions 
that are admitted to the problem class differ 
from the original functions only by an integration constant.
For finite $M$, the action transforms as 
$S[\phi]\mapsto S[\phi]+Mc\phi(0)+\half Mc^2$ under a global translation $\vTh_c$. 
Therefore, the path integral can equally well be defined with an action
given by 
\begin{equation}
   F(\xi[\phi])-\half M\phi(0)^2+S[\phi] \;\;,
\end{equation}
where $\xi$ can be any linear functional such that 
$\xi[\vTh_c\phi]=\xi[\phi]+c$, and $F$ can be any function with the 
restriction that $I[F]$ (\eqn{defI}) exists. 
This results in a replacement of the boundary condition $\phi(0)=0$ by a 
boundary condition dictated by 
$F(\xi[\phi])$, and this is the gauge freedom.

\subsection{The zeroth order contribution}
To calculate the zeroth order contribution to the path integral, the 
eigenfunctions and the corresponding eigenvalues of $A$ have to be found. 
We choose the gauge in which $\xi[\phi]=\intk\phi(x)\,dx$
and $F(\xi[\phi])=\half M\xi^2[\phi]$, so that $A$ with boundary conditions is 
given by
\begin{align}
  &(A\phi)(x)
   \;=\; -\phi^{\pr\pr}(x) - 2z\phi(x) + 2z\intk\phi(y)\,dy \;\;,\\
  &\intk\phi(x)\,dx=0 \quad\textrm{and}\quad \phi^\pr(1)=\phi^\pr(0)=0 \;\;.
\end{align}
The eigenfunctions and the eigenvalues are given by  
\begin{equation}
   u_k(x)
   \;=\; \sqrt{2}\,\cos(k\pi x) \;\;,\quad
         \la_k=k^2\pi^2-2z\;\;,\quad k=1,2,\ldots \;\;.
\end{equation}
According to \eqn{G0}, the zeroth order contribution is given by 
\begin{equation}
   G_0(z)
   \;=\; \left(\prod_{k=1}^{\infty}
         \frac{\la_k(0)}{\la_k(z)}\right)^{\frac{1}{2}}
   \;=\; \left(\prod_{k=1}^{\infty}\frac{k^2\pi^2}{k^2\pi^2-2z}
              \right)^{\frac{1}{2}}
   \;=\; \left(\frac{\sqrt{2z}}{\sin\sqrt{2z}}\right)^{\frac{1}{2}} \;\;,
\end{equation}
and this is the well known expression for the  generating function of the 
probability distribution of the $L_2^*$-discrepancy in one dimension for 
asymptotically large $N$.

In a general gauge, with quadratic $F$, the operator $A$ including the boundary 
conditions is given by 
\begin{align}
   (A\phi)(x)
   \;=\;&  -\phi^{\pr\pr}(x) - 2z\phi(x) + 2z\intk\phi(y)\,dy \nl
        &  +[\de(x-1)-\de(x)]\phi^\pr(x) + M\rho(x)\intk\phi(y)\rho(y)\,dy \;\;,
\end{align}
where $\rho$ is a distribution with the only restriction that it integrates 
to one. Notice that, if $\rho=1$ is taken with $M\ra\infty$, then $A$ in the 
previous gauge is obtained. By integration of the eigenvalue 
equation of $A$ in the general gauge, it is easy to see that 
\begin{equation}
   M\rho(x)\intk\phi(y)\rho(y)\,dy
   \;=\; \la\rho(x)\intk\phi(y)\,dy \;\;,
\end{equation}
where $\la$ is the eigenvalue, so that the eigenvalue equation for $\la=0$ 
becomes 
\begin{equation}
   -\phi^{\pr\pr}(x) - 2z\phi(x) + 2z\intk\phi(y)\,dy
   + [\de(x-1)-\de(x)]\phi^\pr(x) \;=\;0 \;\;.
\end{equation}
The term with the $\de$-distributions just gives the boundary conditions 
$\phi^\pr(0)=\phi^\pr(1)=0$ and we see that solutions only exist for 
$2z=k^2\pi^2$, $k=1,2,\ldots$. These are the values of $2z$ for which 
$\la(z)=0$ and this result is gauge invariant. 
According to \eqn{G0}, they are equal to the  values of $z$ for which 
$1/G_0(z)=0$. With the use of the factor theorem of Weierstrass 
(cf. \cite{whitwat}) we can give a first impulse to calculate $G_0(z)$ in the 
general gauge and write, with $1/G_0(z)^2=f(2z)$,
\begin{equation}
   f(2z)
   \;=\; f(0)e^{\frac{f^\pr(0)}{f(0)}\,2z}
         \prod_{k=0}^\infty\left\{\left(1-\frac{2z}{k^2\pi^2}\right)
	                               e^{\sfrac{2z}{n^2\pi^2}}\right\}
   \;=\; f(0)e^{(\frac{f^\pr(0)}{f(0)}+\frac{1}{6})2z}\,
         \frac{\sin\sqrt{2z}}{\sqrt{2z}} \;\;,
\end{equation}
We have checked in a number of gauges that $f^\prime(0)/f(0)=-\frac{1}{6}$, 
but we did not bother to prove it in the general gauge.

\subsection{The propagator}
We give the propagator $\prop$ in the first gauge of the previous section. 
It has to satisfy the 
equation $(A\prop)(x,y)=\de(x-y)$ including the boundary conditions. If we 
choose the same gauge as in the beginning of the previous section, then the 
propagator has to satisfy, with $u=\sqrt{2z}$,  
\begin{align}
 & \frac{d^2\prop}{dx^2}(x,y) + u^2\prop(x,y) 
   \;=\; -\de(x-y) \;\;,\\
 & \frac{d\prop}{dx}(0,y) = \frac{d\prop}{dx}(1,y) = 0 \;\;,\quad
   \intk\prop(x,y)\,dx=0 \;\;.
\end{align}
The solution can be written down directly in terms of the eigenfunctions 
and the eigenvalues of $A$ and is given by 
\begin{align}
   \prop(x,y) 
  &\;=\; 2\sum_{k=1}^\infty \frac{\cos(k\pi x)\cos(k\pi y)}{k^2\pi^2-u^2}\\
  &\;=\; \frac{1}{u^2} - \frac{1}{2u\sin u}\left\{\cos[u(1-|x+y|)]
	                                 +\cos[u(1-|x-y|)]\right\} \;\;.
\end{align}

\subsection{Instantons\label{Wieninstantons}}
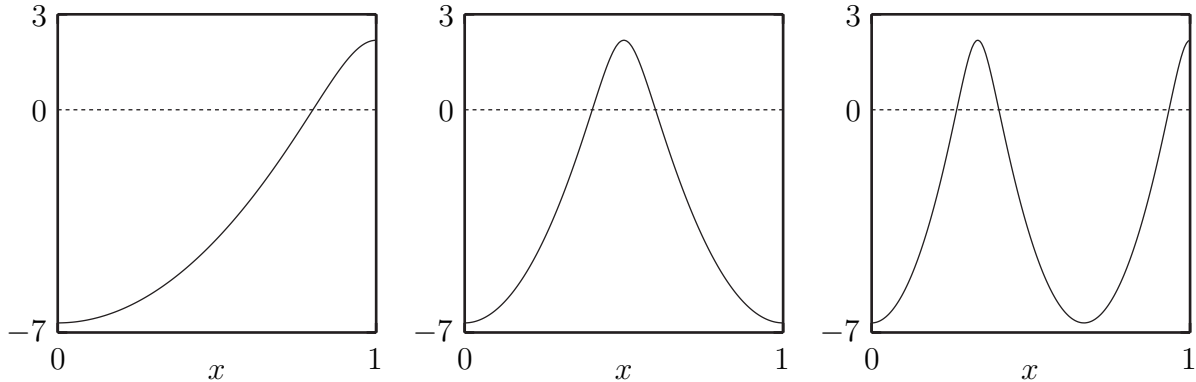
\begin{figure}
\begin{center}
\begin{picture}(120.0,120.0)(23,0)
\LinAxis(  0.0,  0.0)(120.0,  0.0)(1.0,1,3,0,1)
\LinAxis(  0.0,  0.0)(  0.0,120.0)(1,1,-3,0,1)
\LinAxis(  0.0,120.0)(120.0,120.0)(1.0,1,-3,0,1)
\LinAxis(120.0,  0.0)(120.0,120.0)(1,1,3,0,1)
\Text(  0.0,-10)[]{$  0$}
\Text(120.0,-10)[]{$  1$}
\Text(-12,  0.0)[]{$ -7$}
\Text(-7, 84.0)[]{$  0$}
\Text(-7,120.0)[]{$  3$}
\DashLine(0,84)(120,84){1.5}
\Text(60,-15)[]{$x$}
\Curve{(  0.00,  3.61) (  1.20,  3.63) (  2.40,  3.67) (  3.60,  3.73)
       (  4.80,  3.82) (  6.00,  3.93) (  7.20,  4.07) (  8.40,  4.24)
       (  9.60,  4.43) ( 10.80,  4.64) ( 12.00,  4.89) ( 13.20,  5.15)
       ( 14.40,  5.45) ( 15.60,  5.76) ( 16.80,  6.11) ( 18.00,  6.47)
       ( 19.20,  6.87) ( 20.40,  7.29) ( 21.60,  7.73) ( 22.80,  8.20)
       ( 24.00,  8.70) ( 25.20,  9.22) ( 26.40,  9.77) ( 27.60, 10.34)
       ( 28.80, 10.94) ( 30.00, 11.56) ( 31.20, 12.21) ( 32.40, 12.88)
       ( 33.60, 13.58) ( 34.80, 14.30) ( 36.00, 15.05) ( 37.20, 15.83)
       ( 38.40, 16.63) ( 39.60, 17.45) ( 40.80, 18.31) ( 42.00, 19.18)
       ( 43.20, 20.08) ( 44.40, 21.01) ( 45.60, 21.96) ( 46.80, 22.94)
       ( 48.00, 23.94) ( 49.20, 24.97) ( 50.40, 26.03) ( 51.60, 27.11)
       ( 52.80, 28.21) ( 54.00, 29.34) ( 55.20, 30.49) ( 56.40, 31.67)
       ( 57.60, 32.88) ( 58.80, 34.11) ( 60.00, 35.36) ( 61.20, 36.64)
       ( 62.40, 37.95) ( 63.60, 39.28) ( 64.80, 40.63) ( 66.00, 42.01)
       ( 67.20, 43.42) ( 68.40, 44.85) ( 69.60, 46.30) ( 70.80, 47.78)
       ( 72.00, 49.28) ( 73.20, 50.80) ( 74.40, 52.35) ( 75.60, 53.93)
       ( 76.80, 55.52) ( 78.00, 57.14) ( 79.20, 58.79) ( 80.40, 60.45)
       ( 81.60, 62.14) ( 82.80, 63.85) ( 84.00, 65.57) ( 85.20, 67.32)
       ( 86.40, 69.09) ( 87.60, 70.88) ( 88.80, 72.68) ( 90.00, 74.50)
       ( 91.20, 76.33) ( 92.40, 78.17) ( 93.60, 80.03) ( 94.80, 81.89)
       ( 96.00, 83.76) ( 97.20, 85.62) ( 98.40, 87.48) ( 99.60, 89.34)
       (100.80, 91.18) (102.00, 92.99) (103.20, 94.78) (104.40, 96.54)
       (105.60, 98.24) (106.80, 99.89) (108.00,101.47) (109.20,102.96)
       (110.40,104.35) (111.60,105.64) (112.80,106.79) (114.00,107.80)
       (115.20,108.65) (116.40,109.32) (117.60,109.81) (118.80,110.11)
       (120.00,110.21)}
\end{picture}
\begin{picture}(120.0,120.0)(-7,0)
\LinAxis(  0.0,  0.0)(120.0,  0.0)(1.0,1,3,0,1)
\LinAxis(  0.0,  0.0)(  0.0,120.0)(1,1,-3,0,1)
\LinAxis(  0.0,120.0)(120.0,120.0)(1.0,1,-3,0,1)
\LinAxis(120.0,  0.0)(120.0,120.0)(1,1,3,0,1)
\Text(  0.0,-10)[]{$  0$}
\Text(120.0,-10)[]{$  1$}
\Text(-12,  0.0)[]{$ -7$}
\Text(-7, 84.0)[]{$  0$}
\Text(-7,120.0)[]{$  3$}
\DashLine(0,84)(120,84){1.5}
\Text(60,-15)[]{$x$}
\Curve{(  0.00,  3.61) (  1.20,  3.67) (  2.40,  3.82) (  3.60,  4.07)
       (  4.80,  4.43) (  6.00,  4.89) (  7.20,  5.45) (  8.40,  6.11)
       (  9.60,  6.87) ( 10.80,  7.73) ( 12.00,  8.70) ( 13.20,  9.77)
       ( 14.40, 10.94) ( 15.60, 12.21) ( 16.80, 13.58) ( 18.00, 15.05)
       ( 19.20, 16.63) ( 20.40, 18.31) ( 21.60, 20.08) ( 22.80, 21.96)
       ( 24.00, 23.94) ( 25.20, 26.03) ( 26.40, 28.21) ( 27.60, 30.49)
       ( 28.80, 32.88) ( 30.00, 35.36) ( 31.20, 37.95) ( 32.40, 40.63)
       ( 33.60, 43.42) ( 34.80, 46.30) ( 36.00, 49.28) ( 37.20, 52.35)
       ( 38.40, 55.52) ( 39.60, 58.79) ( 40.80, 62.14) ( 42.00, 65.57)
       ( 43.20, 69.09) ( 44.40, 72.68) ( 45.60, 76.33) ( 46.80, 80.03)
       ( 48.00, 83.76) ( 49.20, 87.48) ( 50.40, 91.18) ( 51.60, 94.78)
       ( 52.80, 98.24) ( 54.00,101.47) ( 55.20,104.35) ( 56.40,106.79)
       ( 57.60,108.65) ( 58.80,109.81) ( 60.00,110.21) ( 61.20,109.81)
       ( 62.40,108.65) ( 63.60,106.79) ( 64.80,104.35) ( 66.00,101.47)
       ( 67.20, 98.24) ( 68.40, 94.78) ( 69.60, 91.18) ( 70.80, 87.48)
       ( 72.00, 83.76) ( 73.20, 80.03) ( 74.40, 76.33) ( 75.60, 72.68)
       ( 76.80, 69.09) ( 78.00, 65.57) ( 79.20, 62.14) ( 80.40, 58.79)
       ( 81.60, 55.52) ( 82.80, 52.35) ( 84.00, 49.28) ( 85.20, 46.30)
       ( 86.40, 43.42) ( 87.60, 40.63) ( 88.80, 37.95) ( 90.00, 35.36)
       ( 91.20, 32.88) ( 92.40, 30.49) ( 93.60, 28.21) ( 94.80, 26.03)
       ( 96.00, 23.94) ( 97.20, 21.96) ( 98.40, 20.08) ( 99.60, 18.31)
       (100.80, 16.63) (102.00, 15.05) (103.20, 13.58) (104.40, 12.21)
       (105.60, 10.94) (106.80,  9.77) (108.00,  8.70) (109.20,  7.73)
       (110.40,  6.87) (111.60,  6.11) (112.80,  5.45) (114.00,  4.89)
       (115.20,  4.43) (116.40,  4.07) (117.60,  3.82) (118.80,  3.67)
       (120.00,  3.61)}
\end{picture}
\begin{picture}(120.0,120.0)(-37,0)
\LinAxis(  0.0,  0.0)(120.0,  0.0)(1.0,1,3,0,1)
\LinAxis(  0.0,  0.0)(  0.0,120.0)(1,1,-3,0,1)
\LinAxis(  0.0,120.0)(120.0,120.0)(1.0,1,-3,0,1)
\LinAxis(120.0,  0.0)(120.0,120.0)(1,1,3,0,1)
\Text(  0.0,-10)[]{$  0$}
\Text(120.0,-10)[]{$  1$}
\Text(-12,  0.0)[]{$ -7$}
\Text(-7, 84.0)[]{$  0$}
\Text(-7,120.0)[]{$  3$}
\DashLine(0,84)(120,84){1.5}
\Text(60,-15)[]{$x$}
\Curve{(  0.00,  3.61) (  1.20,  3.73) (  2.40,  4.07) (  3.60,  4.64)
       (  4.80,  5.45) (  6.00,  6.47) (  7.20,  7.73) (  8.40,  9.22)
       (  9.60, 10.94) ( 10.80, 12.88) ( 12.00, 15.05) ( 13.20, 17.45)
       ( 14.40, 20.08) ( 15.60, 22.94) ( 16.80, 26.03) ( 18.00, 29.34)
       ( 19.20, 32.88) ( 20.40, 36.64) ( 21.60, 40.63) ( 22.80, 44.85)
       ( 24.00, 49.28) ( 25.20, 53.93) ( 26.40, 58.79) ( 27.60, 63.85)
       ( 28.80, 69.09) ( 30.00, 74.50) ( 31.20, 80.03) ( 32.40, 85.62)
       ( 33.60, 91.18) ( 34.80, 96.54) ( 36.00,101.47) ( 37.20,105.64)
       ( 38.40,108.65) ( 39.60,110.11) ( 41.20,109.32) ( 42.40,106.79)
       ( 43.60,102.96) ( 44.80, 98.24) ( 46.00, 92.99) ( 47.20, 87.48)
       ( 48.40, 81.89) ( 49.60, 76.33) ( 50.80, 70.88) ( 52.00, 65.57)
       ( 53.20, 60.45) ( 54.40, 55.52) ( 55.60, 50.80) ( 56.80, 46.30)
       ( 58.00, 42.01) ( 59.20, 37.95) ( 60.40, 34.11) ( 61.60, 30.49)
       ( 62.80, 27.11) ( 64.00, 23.94) ( 65.20, 21.01) ( 66.40, 18.31)
       ( 67.60, 15.83) ( 68.80, 13.58) ( 70.00, 11.56) ( 71.20,  9.77)
       ( 72.40,  8.20) ( 73.60,  6.87) ( 74.80,  5.76) ( 76.00,  4.89)
       ( 77.20,  4.24) ( 78.40,  3.82) ( 79.60,  3.63) ( 81.20,  3.73)
       ( 82.40,  4.07) ( 83.60,  4.64) ( 84.80,  5.45) ( 86.00,  6.47)
       ( 87.20,  7.73) ( 88.40,  9.22) ( 89.60, 10.94) ( 90.80, 12.88)
       ( 92.00, 15.05) ( 93.20, 17.45) ( 94.40, 20.08) ( 95.60, 22.94)
       ( 96.80, 26.03) ( 98.00, 29.34) ( 99.20, 32.88) (100.40, 36.64)
       (101.60, 40.63) (102.80, 44.85) (104.00, 49.28) (105.20, 53.93)
       (106.40, 58.79) (107.60, 63.85) (108.80, 69.09) (110.00, 74.50)
       (111.20, 80.03) (112.40, 85.62) (113.60, 91.18) (114.80, 96.54)
       (116.00,101.47) (117.20,105.64) (118.40,108.65) (119.60,110.11)}
\end{picture}
\vspace{20pt}
\caption{Instanton solutions $\phi_k$ with $E=5.7$ and number of 
         bending points $k=1,2,3$.}
\label{figinstant}	 
\end{center}
\end{figure}

To start, we repeat that non-trivial  
instanton solutions only exist if $z\in[0,\infty)$ (section \ref{geninst}).
In order to investigate the instantons in the Wiener problem class, we analyze 
$\Si[\phi]=S[\phi/g]/N$, because this new action does not depend on $N$: 
\begin{equation}
   \Si[\phi]
   \;=\; \frac{1}{4z}\intk \phi^\pr(x)^2\,dx + \intk \phi(x)\,dx 
         - \log\left(\intk e^{\phi(x)}\,dx\right) \;\;.
\end{equation}
Extremal points of this action are solutions of the field equation 
\begin{equation}
   -\frac{1}{2z}\,\phi^{\pr\pr}(x) + 1 
   - \frac{e^{\phi(x)}}{\intk e^{\phi(y)}\,dy}
   \;=\; 0 \;\;
\end{equation}
that also satisfy the boundary conditions, for which we take 
$\phi(0)=\phi^\pr(1)=0$ at this point. Because the action as well as the 
equation is 
invariant under global translations, 
solutions can always be chosen such that $\intk\exp(\phi(y))\,dy=1$, so that 
the equation becomes
\begin{equation}
   -\frac{1}{2z}\,\phi^{\pr\pr}(x) + 1 - e^{\phi(x)} \;=\; 0 
   \;\;,\quad\textrm{with}\;\;
   \phi^\pr(1)=0\;\;\textrm{and}\;\;\intk e^{\phi(x)}\,dx=1\;\;.
\end{equation}
We must also have $\phi^\pr(0)=0$. 
The problem is now reduced to that of the motion of a classical 
particle with a mass $1/\sqrt{4z}$\, in 
a potential 
\begin{equation}
   U(\phi) \;=\; e^\phi-\phi-1 
\end{equation}
and the solution can be written implicitly as 
\begin{equation}
   \sqrt{4z}\,\frac{dx}{d\phi} \;=\; \frac{1}{\sqrt{E-U(\phi)}} \;\;,
\label{implicit}   
\end{equation}
where the integration constant $E$, the {\em energy}, has to be larger than 
zero for solutions to exist. It is easy to see that the solutions are 
oscillatory and that, if $\phi(x)$ is a solution with one bending point, then 
also 
\begin{equation}
   \phi_k(x) \;=\; 
   \begin{cases}
      \phi(kx-p)   & \frac{p}{k}\leq x \leq\frac{p+1}{k}\;\;\textrm{$p$ even}\\
      \phi(1+p-kx) & \frac{p}{k}\leq x \leq\frac{p+1}{k}\;\;\textrm{$p$ odd}
   \end{cases}
   \;,\quad p = 0,1,\ldots,k-1\;\;,
\end{equation}
is a solution for $k=2,3,\ldots$.
These new solutions have the same energy, but a larger number of bending 
points, namely $k$, and the value of $z$ increases by a factor $k^2$. 
Hence we can classify the solutions 
according to the energy and the number bending points.  
This classification in terms of the number of bending points is quite natural 
and this can best be understood by looking at the limit of $N\ra\infty$. 
Then, the equation becomes 
\begin{equation}
   -\phi^{\pr\pr}(x)-2z\phi(x)+2z\intk\phi(y)\,dy \;=\; 0 \;\;,
\end{equation}
with $\phi(0)=\phi^\pr(1)=0$ and the solutions are given by 
\begin{equation}
   \phi_k(x)=\textstyle{\sqrt{\frac{2}{3}}}\,[1-\cos(k\pi x)]\;\;,\;\;
   2z=k^2\pi^2\;\;,\;\;
   k=1,2,\ldots \;\;,
\end{equation}
so that the instantons are completely classified with the number of bending 
points $k$. 
If $N$ becomes finite, these solutions are deformed but keep the same value 
of $k$ (\fig{figinstant}). For given $k$ there are infinitely many solutions 
classified by $E$. 

We now concentrate on the instantons with one bending point, because 
the numerical value of the action is independent of the number of bending 
points. 
Those instantons are completely characterized by their energy. The values of 
$z$ for which these instantons exist are defined as a function of $E$ by 
\eqn{implicit}, which states that 
\begin{equation}
   T(E)
   \;\equiv\; \sqrt{4z}
   \;=\; \int\limits_{\phi_-}^{\phi_+}\frac{d\phi}{\sqrt{E-U(\phi)}} \;\;,
\label{period}   
\end{equation}
where $\phi_-$ and $\phi_+$ are the classical turning points. 
They are solutions of $U(\phi_\pm)=E$ with $\phi_-<0<\phi_+$.
In classical mechanics, $T(E)$ is proportional to the period of 
a particle in the potential $U$ (cf. \cite{LandL}).

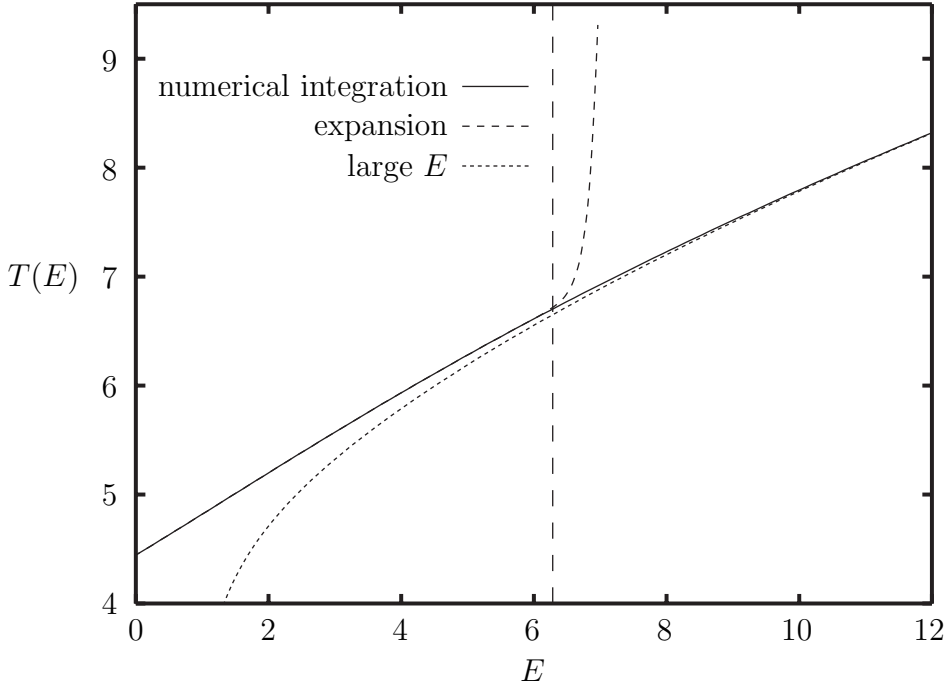
\begin{figure}[t]
\begin{center}
\begin{picture}(300,226)(-15,0)
\LinAxis(0,0)(300,0)(6,1,3,0,1.5)
\LinAxis(0,0)(0,226)(5.5,1,-3,0,1.5)
\LinAxis(0,226)(300,226)(6,1,-3,0,1.5)
\LinAxis(300,0)(300,226)(5.5,1,3,0,1.5)
\Text(118,195)[r]{numerical integration}\Line(123,195)(148,195)
\Text(118,180)[r]{expansion}\DashLine(123,180)(148,180){3}
\Text(118,165)[r]{large $E$}\DashLine(123,165)(148,165){1.5}
\Text(0,-10)[]{$0$}\Text(50,-10)[]{$2$}\Text(100,-10)[]{$4$}
\Text(150,-10)[]{$6$}\Text(200,-10)[]{$8$}\Text(250,-10)[]{$10$}
\Text(300,-10)[]{$12$}\Text(-10,0)[]{$4$}\Text(-10,41.1)[]{$5$}
\Text(-10,82.2)[]{$6$}\Text(-10,123.3)[]{$7$}
\Text(-10,164.4)[]{$8$}\Text(-10,205.5)[]{$9$}
\Text(150,-25)[]{$E$}\Text(-35,123.3)[]{$T(E)$}
\DashLine(157.1,0)(157.1,226){6}
\Curve{(  0.2, 18.4) (  3.2, 20.2) (  6.2, 22.0) (  9.2, 23.9) ( 12.2, 25.7)
       ( 15.2, 27.6) ( 18.2, 29.4) ( 21.2, 31.3) ( 24.2, 33.2) ( 27.2, 35.0)
       ( 30.2, 36.9) ( 33.2, 38.8) ( 36.2, 40.7) ( 39.2, 42.5) ( 42.2, 44.4)
       ( 45.2, 46.3) ( 48.2, 48.1) ( 51.2, 50.0) ( 54.2, 51.9) ( 57.2, 53.7)
       ( 60.2, 55.6) ( 63.2, 57.4) ( 66.2, 59.2) ( 69.2, 61.1) ( 72.2, 62.9)
       ( 75.2, 64.7) ( 78.2, 66.5) ( 81.2, 68.3) ( 84.2, 70.1) ( 87.2, 71.9)
       ( 90.2, 73.7) ( 93.2, 75.5) ( 96.2, 77.2) ( 99.2, 79.0) (102.2, 80.7)
       (105.2, 82.5) (108.2, 84.2) (111.2, 85.9) (114.2, 87.6) (117.2, 89.3)
       (120.2, 91.0) (123.1, 92.7) (126.1, 94.3) (129.1, 96.0) (132.1, 97.7)
       (135.1, 99.3) (138.1,100.9) (141.1,102.6) (144.1,104.2) (147.1,105.8)
       (150.1,107.4) (153.1,109.0) (156.1,110.5) (159.1,112.1) (162.1,113.7)
       (165.1,115.2) (168.1,116.8) (171.1,118.3) (174.1,119.8) (177.1,121.3)
       (180.1,122.8) (183.1,124.3) (186.1,125.8) (189.1,127.3) (192.1,128.8)
       (195.1,130.3) (198.1,131.7) (201.1,133.2) (204.1,134.6) (207.1,136.1)
       (210.1,137.5) (213.1,138.9) (216.1,140.4) (219.1,141.8) (222.1,143.2)
       (225.1,144.6) (228.1,146.0) (231.1,147.3) (234.1,148.7) (237.1,150.1)
       (240.1,151.5) (243.0,152.8) (246.0,154.2) (249.0,155.5) (252.0,156.8)
       (255.0,158.2) (258.0,159.5) (261.0,160.8) (264.0,162.1) (267.0,163.5)
       (270.0,164.8) (273.0,166.1) (276.0,167.4) (279.0,168.6) (282.0,169.9)
       (285.0,171.2) (288.0,172.5) (291.0,173.7) (294.0,175.0) (297.0,176.3)
       (300.0,177.5) }
\DashCurve{(  0.2, 18.4) (  3.2, 20.2) (  6.2, 22.0) (  9.2, 23.9) ( 12.2, 25.7)
       ( 15.2, 27.6) ( 18.2, 29.4) ( 21.2, 31.3) ( 24.2, 33.2) ( 27.2, 35.0)
       ( 30.2, 36.9) ( 33.2, 38.8) ( 36.2, 40.7) ( 39.2, 42.5) ( 42.2, 44.4)
       ( 45.2, 46.3) ( 48.2, 48.1) ( 51.2, 50.0) ( 54.2, 51.9) ( 57.2, 53.7)
       ( 60.2, 55.6) ( 63.2, 57.4) ( 66.2, 59.2) ( 69.2, 61.1) ( 72.2, 62.9)
       ( 75.2, 64.7) ( 78.2, 66.5) ( 81.2, 68.3) ( 84.2, 70.1) ( 87.2, 71.9)
       ( 90.2, 73.7) ( 93.2, 75.5) ( 96.2, 77.2) ( 99.2, 79.0) (102.2, 80.7)
       (105.2, 82.5) (108.2, 84.2) (111.2, 85.9) (114.2, 87.6) (117.2, 89.3)
       (120.2, 91.0) (123.1, 92.7) (126.1, 94.3) (129.1, 96.0) (132.1, 97.7)
       (135.1, 99.3) (138.1,100.9) (141.1,102.6) (144.1,104.2) (147.1,105.8)
       (150.1,107.4) (153.1,109.1) (156.1,111.0) (159.1,113.2) (162.1,116.4)
       (165.1,122.2) (168.1,133.8) (171.1,159.6) (174.1,218.2) }{3}
\DashCurve{( 33.2,  0.0) ( 36.2,  7.1) ( 39.2, 13.0) ( 42.2, 18.2) ( 45.2, 22.8)
       ( 48.2, 26.9) ( 51.2, 30.7) ( 54.2, 34.2) ( 57.2, 37.6) ( 60.2, 40.7)
       ( 63.2, 43.7) ( 66.2, 46.6) ( 69.2, 49.3) ( 72.2, 52.0) ( 75.2, 54.5)
       ( 78.2, 57.0) ( 81.2, 59.5) ( 84.2, 61.8) ( 87.2, 64.1) ( 90.2, 66.4)
       ( 93.2, 68.6) ( 96.2, 70.8) ( 99.2, 72.9) (102.2, 75.0) (105.2, 77.1)
       (108.2, 79.1) (111.2, 81.1) (114.2, 83.1) (117.2, 85.0) (120.2, 86.9)
       (123.1, 88.8) (126.1, 90.7) (129.1, 92.6) (132.1, 94.4) (135.1, 96.2)
       (138.1, 98.0) (141.1, 99.8) (144.1,101.5) (147.1,103.2) (150.1,105.0)
       (153.1,106.7) (156.1,108.3) (159.1,110.0) (162.1,111.7) (165.1,113.3)
       (168.1,115.0) (171.1,116.6) (174.1,118.2) (177.1,119.8) (180.1,121.4)
       (183.1,122.9) (186.1,124.5) (189.1,126.0) (192.1,127.6) (195.1,129.1)
       (198.1,130.6) (201.1,132.1) (204.1,133.6) (207.1,135.1) (210.1,136.6)
       (213.1,138.0) (216.1,139.5) (219.1,141.0) (222.1,142.4) (225.1,143.8)
       (228.1,145.3) (231.1,146.7) (234.1,148.1) (237.1,149.5) (240.1,150.9)
       (243.0,152.3) (246.0,153.7) (249.0,155.0) (252.0,156.4) (255.0,157.7)
       (258.0,159.1) (261.0,160.4) (264.0,161.8) (267.0,163.1) (270.0,164.4)
       (273.0,165.8) (276.0,167.1) (279.0,168.4) (282.0,169.7) (285.0,171.0)
       (288.0,172.3) (291.0,173.5) (294.0,174.8) (297.0,176.1) (300.0,177.4) 
       }{1.5}
\end{picture} 
\vspace{25pt}
\caption{$T(E)$ computed by numerical integration, as an expansion around $E=0$
         and as an approximation for large $E$. The expansion is up to and 
	 including $\Ord(E^{49})$.}
\label{figT(E)}
\end{center}
\end{figure}

The function $T$ cannot be expressed in terms of elementary functions, but a 
number of its properties can be derived, as we shall now discuss.
For small $E$, a quadratic approximation of the potential can be made with 
$\phi_\pm=\pm\sqrt{2E}$ with the result that 
\begin{equation}
   \lim_{E\da0}T(E)
   \;=\; \pi\sqrt{2}  \quad\Longrightarrow\quad
   \lim_{E\da0}z(E)
   \;=\; \half\pi^2  \;\;.
\end{equation}
The question is now whether $z$ is 
increasing as a function of $E$. 
To calculate $T(E)$ for large $E$, $U(\phi)$ can be approximated by 
$-1-\phi$ for $\phi<0$ and by $e^\phi$ for $\phi>0$, so that 
\begin{equation}
   T(E\ra\infty)
   \;\approx\; 2\sqrt{E+1} 
         + \frac{2}{\sqrt{E}}\,\log\left(\sqrt{E}+\sqrt{E-1}\right) \;\;,
\label{TlargeE}	 
\end{equation}
so $T(E)$ is clearly increasing for large $E$.
To analyze $T(E)$ for small $E$, we make an 
expansion in powers of $E$. Therefore, we write 
\begin{equation}
   T(E)
   \;=\; \int\limits_0^{\sqrt{2E}}\left(E-\half v^2\right)^{-1/2}
         \frac{d}{dv}\left[f(v)-f(-v)\right]\,dv \;\;,
\label{Tintoverv}	 
\end{equation}
where $f$ is a continuous solution of the implicit equation 
\begin{equation}
   e^{f(v)} - f(v) - 1 = \half v^2 \;\;,
\end{equation}
with $f(v)\sim v$ for small $v$. In \cite{hak} it is shown that it is given by 
the function values on the principal Riemann sheet of the general continuous 
solution and that is has an expansion $f(v)=\sum_{n=0}^\infty\al_nv^n$ 
with the coefficients $\al_n$ given by 
\begin{equation}
   \al_1=1\quad\textrm{and}\quad
   \al_n =-\frac{1}{n+1}\left[
          \half(n-1)\al_{n-1}+\sum_{k=2}^{n-1}k\al_k\al_{n+1-k}\right]\;\;
   \textrm{for}\;\; n>1 \;\;,
\end{equation}
and with the radius of convergence equal to $\sqrt{4\pi}$. If we substitute 
the power series into \eqn{Tintoverv} and integrate term by term, we obtain 
the following power series for $|E|<2\pi$:
\begin{equation}
   T(E)
   \;=\; \sum_{n=1,~\textrm{$n$ odd}}^\infty
         \frac{\Gamma(\half)\Gamma(\frac{n}{2})}{\Gamma(\frac{n+1}{2})}\,
	 n\al_n\, 2^\frac{n}{2} E^\frac{n-1}{2} \;\;.
\end{equation}
The first few terms in this expansion are 
\begin{equation}
   T(E) 
   \;=\; \pi\sqrt{2}\left[1 + \frac{E}{12}
                            +\frac{1}{4}\left(\frac{E}{12}\right)^2
			    -\frac{139}{180}\left(\frac{E}{12}\right)^3
			    -\frac{571}{2880}\left(\frac{E}{12}\right)^4
			    +\Ord(E^5)\right] \;\;.
\end{equation}
The asymptotic behavior of the coefficients $\al_n$ has also been determined in 
\cite{hak}, with the result that, for large and integer $k$,  
\begin{equation}
  \al_n \;\sim\; \frac{1}{(4\pi)^\frac{n}{2}\,n^\frac{3}{2}}\times
  \begin{cases}
      -2(-)^k         & \textrm{if $n=4k$}\\
      0               & \textrm{if $n=4k+1$} \\
      -2(-)^k         & \textrm{if $n=4k+2$}\\
      +2\sqrt{2}(-)^k & \textrm{if $n=4k+3$}
  \end{cases}    
\end{equation}
The results are summarized in \fig{figT(E)}. Depicted are the behavior for 
large $E$, the expansion for small $E$ and a numerical evaluation of the 
integral of \eqn{period}. Notice the strong deviation of the expansion from the 
other curves for $E>2\pi$, the radius of convergence. For this plot the first  
$50$ terms were used. It appears that $T$ is indeed an increasing function 
of $E$.

We now turn to the analysis of the value of the action for an instanton. 
In the foregoing, we have shown for which positive values of $z$ no instantons 
exist. Now we will show that the action indeed becomes negative for $z$ 
positive and large enough. For an instanton solution with one bending point,
the action is given by 
\begin{align}
  &S(E)
   \;=\; \frac{1}{4z(E)}\int_0^1\phi^{\pr\pr}(x)\,dx + \int_0^1\phi(x)\,dx
   \;=\; E + 2\,\frac{T_1(E)}{T(E)} \;\;,\label{SofTT1}\\
  &T_1(E)
  \;=\; \int\limits_{\phi_-}^{\phi_+}\frac{\phi\,d\phi}{\sqrt{E-U(\phi)}} 
        \label{defT1}\;\;.
\end{align}
With the use of the same approximations for $U(\phi)$ as in the derivation of 
\eqn{TlargeE}, it is easy to see that, for large $E$, $T_1(E)$ is bounded by  
\begin{equation}
   - \frac{4}{3}(E+1)^{3/2} 
         + \frac{2\log E}{\sqrt{E}}\,\log\left(\sqrt{E}+\sqrt{E-1}\right) \;\;,
	 \label{SlargeE}
\end{equation}
so that $S(E)$ clearly becomes negative for large $E$.

\begin{figure}[t]
\begin{center}
\begin{picture}(300.0,226.0)(-15,0)
\LinAxis(  0.0,  0.0)(300.0,  0.0)(6.0,1,3,0,1.5)
\LinAxis(  0.0,  0.0)(  0.0,226.0)(7.0,1,-3,0,1.5)
\LinAxis(  0.0,226.0)(300.0,226.0)(6.0,1,-3,0,1.5)
\LinAxis(300.0,  0.0)(300.0,226.0)(7.0,1,3,0,1.5)
\Text(  0.0,-10)[]{$  0$}
\Text( 50.0,-10)[]{$  2$}
\Text(100.0,-10)[]{$  4$}
\Text(150.0,-10)[]{$  6$}
\Text(200.0,-10)[]{$  8$}
\Text(250.0,-10)[]{$ 10$}
\Text(300.0,-10)[]{$ 12$}
\Text(-20,  0.0)[]{$ -3.0$}
\Text(-20, 32.3)[]{$ -2.5$}
\Text(-20, 64.6)[]{$ -2.0$}
\Text(-20, 96.9)[]{$ -1.5$}
\Text(-20,129.1)[]{$ -1.0$}
\Text(-20,161.4)[]{$ -0.5$}
\Text(-17,193.7)[]{$  0.0$}
\Text(-17,226.0)[]{$  0.5$}
\Text(150,-25)[]{$E$}
\Text(-50,113)[]{$S(E)$}
\Text(118,65)[r]{numerical integration}\Line(123,65)(148,65)
\Text(118,50)[r]{expansion}\DashLine(123,50)(148,50){3}
\Text(118,35)[r]{large $E$}\DashLine(123,35)(148,35){1.5}
\Line(0,193.7)(300,193.7)
\DashLine(157.1,0)(157.1,226){6.0}
\Curve{(  0.2,193.7) (  3.2,193.7) (  6.2,193.5) (  9.2,193.4) ( 12.2,193.1)
       ( 15.2,192.8) ( 18.2,192.3) ( 21.2,191.9) ( 24.2,191.3) ( 27.2,190.7)
       ( 30.2,190.1) ( 33.2,189.4) ( 36.2,188.6) ( 39.2,187.7) ( 42.2,186.8)
       ( 45.2,185.9) ( 48.2,184.9) ( 51.2,183.9) ( 54.2,182.8) ( 57.2,181.7)
       ( 60.2,180.5) ( 63.2,179.3) ( 66.2,178.0) ( 69.2,176.7) ( 72.2,175.4)
       ( 75.2,174.0) ( 78.2,172.6) ( 81.2,171.2) ( 84.2,169.7) ( 87.2,168.2)
       ( 90.2,166.7) ( 93.2,165.2) ( 96.2,163.6) ( 99.2,162.0) (102.2,160.4)
       (105.2,158.7) (108.2,157.1) (111.2,155.4) (114.2,153.7) (117.2,151.9)
       (120.2,150.2) (123.1,148.4) (126.1,146.6) (129.1,144.8) (132.1,143.0)
       (135.1,141.2) (138.1,139.3) (141.1,137.5) (144.1,135.6) (147.1,133.7)
       (150.1,131.8) (153.1,129.9) (156.1,128.0) (159.1,126.0) (162.1,124.1)
       (165.1,122.1) (168.1,120.2) (171.1,118.2) (174.1,116.2) (177.1,114.2)
       (180.1,112.2) (183.1,110.2) (186.1,108.1) (189.1,106.1) (192.1,104.1)
       (195.1,102.0) (198.1,100.0) (201.1, 97.9) (204.1, 95.8) (207.1, 93.8)
       (210.1, 91.7) (213.1, 89.6) (216.1, 87.5) (219.1, 85.4) (222.1, 83.3)
       (225.1, 81.2) (228.1, 79.1) (231.1, 76.9) (234.1, 74.8) (237.1, 72.7)
       (240.1, 70.5) (243.0, 68.4) (246.0, 66.3) (249.0, 64.1) (252.0, 61.9)
       (255.0, 59.8) (258.0, 57.6) (261.0, 55.4) (264.0, 53.3) (267.0, 51.1)
       (270.0, 48.9) (273.0, 46.7) (276.0, 44.5) (279.0, 42.3) (282.0, 40.1)
       (285.0, 37.9) (288.0, 35.7) (291.0, 33.5) (294.0, 31.3) (297.0, 29.1)
       (300.0, 26.9)}
\DashCurve{(  0.2,193.7) (  3.2,193.7) (  6.2,193.5) (  9.2,193.4) ( 12.2,193.1)
       ( 15.2,192.8) ( 18.2,192.3) ( 21.2,191.9) ( 24.2,191.3) ( 27.2,190.7)
       ( 30.2,190.1) ( 33.2,189.3) ( 36.2,188.6) ( 39.2,187.7) ( 42.2,186.8)
       ( 45.2,185.9) ( 48.2,184.9) ( 51.2,183.9) ( 54.2,182.8) ( 57.2,181.7)
       ( 60.2,180.5) ( 63.2,179.3) ( 66.2,178.0) ( 69.2,176.7) ( 72.2,175.4)
       ( 75.2,174.0) ( 78.2,172.6) ( 81.2,171.2) ( 84.2,169.7) ( 87.2,168.2)
       ( 90.2,166.7) ( 93.2,165.2) ( 96.2,163.6) ( 99.2,162.0) (102.2,160.4)
       (105.2,158.7) (108.2,157.1) (111.2,155.4) (114.2,153.6) (117.2,151.9)
       (120.2,150.2) (123.1,148.4) (126.1,146.6) (129.1,144.8) (132.1,143.0)
       (135.1,141.2) (138.1,139.3) (141.1,137.5) (144.1,135.6) (147.1,133.7)
       (150.1,131.9) (153.1,130.1) (156.1,128.6) (159.1,127.5) (162.1,127.8)
       (165.1,131.1) (168.1,141.5) (171.1,168.0) (173.9,226.0)}{3.0}
\DashCurve{( 27.2,115.3) ( 30.2,127.3) ( 33.2,135.2) ( 36.2,141.1) ( 39.2,145.8)
       ( 42.2,149.6) ( 45.2,152.7) ( 48.2,155.3) ( 51.2,157.4) ( 54.2,159.1)
       ( 57.2,160.5) ( 60.2,161.7) ( 63.2,162.5) ( 66.2,163.1) ( 69.2,163.6)
       ( 72.2,163.8) ( 75.2,163.9) ( 78.2,163.9) ( 81.2,163.7) ( 84.2,163.4)
       ( 87.2,163.0) ( 90.2,162.5) ( 93.2,161.9) ( 96.2,161.2) ( 99.2,160.4)
       (102.2,159.6) (105.2,158.6) (108.2,157.7) (111.2,156.6) (114.2,155.5)
       (117.2,154.4) (120.2,153.2) (123.1,151.9) (126.1,150.6) (129.1,149.3)
       (132.1,147.9) (135.1,146.5) (138.1,145.0) (141.1,143.6) (144.1,142.0)
       (147.1,140.5) (150.1,138.9) (153.1,137.3) (156.1,135.7) (159.1,134.1)
       (162.1,132.4) (165.1,130.7) (168.1,129.0) (171.1,127.2) (174.1,125.5)
       (177.1,123.7) (180.1,121.9) (183.1,120.1) (186.1,118.3) (189.1,116.4)
       (192.1,114.6) (195.1,112.7) (198.1,110.8) (201.1,108.9) (204.1,107.0)
       (207.1,105.1) (210.1,103.1) (213.1,101.2) (216.1, 99.2) (219.1, 97.2)
       (222.1, 95.2) (225.1, 93.2) (228.1, 91.2) (231.1, 89.2) (234.1, 87.2)
       (237.1, 85.1) (240.1, 83.1) (243.0, 81.1) (246.0, 79.0) (249.0, 76.9)
       (252.0, 74.8) (255.0, 72.8) (258.0, 70.7) (261.0, 68.6) (264.0, 66.5)
       (267.0, 64.3) (270.0, 62.2) (273.0, 60.1) (276.0, 58.0) (279.0, 55.8)
       (282.0, 53.7) (285.0, 51.5) (288.0, 49.4) (291.0, 47.2) (294.0, 45.1)
       (297.0, 42.9) (300.0, 40.7)}{1.5}
\end{picture}
\vspace{25pt}
\caption{$S(E)$ computed by numerical integration and as an expansion around 
         $E=0$. The expansion is up to and including $\Ord(E^{48})$ and its 
	 radius of convergence is $2\pi$.
	 The curve for large $E$ is the upper bound of \eqn{SlargeE}.} 
\label{figS(E)}
\end{center}
\end{figure}
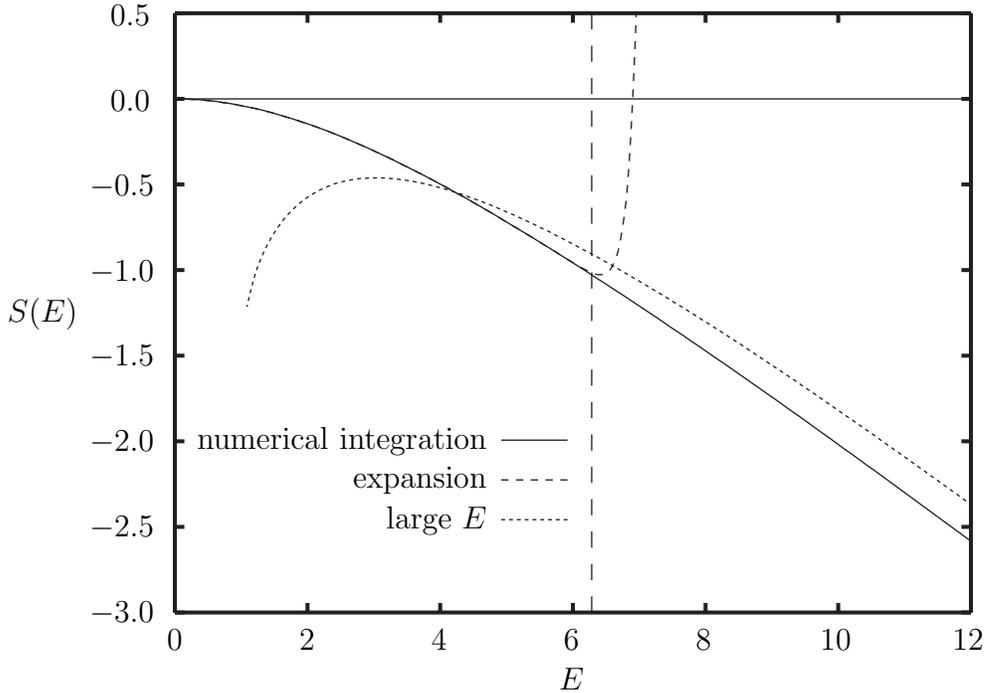
To investigate the behavior of $S(E)$ for small $E$, we use an expansion again.
It can be obtained using \eqn{SofTT1} and the relation 
\begin{equation}
   \frac{dT_1}{dE}(E)
   \;=\; -\frac{1}{2}\,T(E) - E\,\frac{dT}{dE}(E) \;\;.
\label{diffTT1}   
\end{equation}
A derivation of this relation is given in Appendix C. For $E\da0$ a 
quadratic approximation of the potential $U(\phi)$ can be used in \eqn{defT1} 
and we find that $T_1(0)=0$, so that the expansion of $T(E)$ can be substituted 
in \eqn{diffTT1} and an expansion of $T_1(E)$ can be obtained by integrating 
term by term. The expansions of $T(E)$ and $T_1(E)$ can then be used to find 
the expansion of $S(E)$ using \eqn{SofTT1}. The first few terms are 
\begin{align}
  &T_1(E)
  \;=\; \pi\sqrt{2}\left[-\frac{E}{2}-\frac{E^2}{16}-\frac{5\,E^3}{3456}
                         +\frac{973\,E^4}{2488320}+\Ord(E^5)\right]  \;\;,\\
  &S(E) 
  \;=\; -\frac{E^2}{24}+\frac{E^3}{432}+\frac{89\,E^4}{414720}+\Ord(E^5) \;\;.
\end{align}
In \fig{figS(E)}, we plot $S(E)$ as obtained from the series expansion, from 
the asymptotic behavior, and from numerical integration. The conclusion is 
that $S(E)$ is always negative.

\section{Conclusion}
We have introduced the machinery of QFT to calculate the moment generating 
function $G$ of the probability distribution $H$ under sets of random points 
of a quadratic discrepancy $D_N$ as a perturbation series on the 
generating function $G_0$ of the distribution $H_0$  
for asymptotically large number of points $N$. We used the fact that $D_N$ 
can be defined as an average-case complexity over a function class and 
presented the formula for $G$ itself as an average over that function class. 
We interpreted this formula as a Euclidean path integral and introduced the 
saddle point approximation to generate a perturbation series in $1/N$. 
This series can 
be seen as a diagrammatic expansion with a propagator $\prop$, which we have 
shown to possess a gauge freedom. Furthermore, we have addressed
the problem of phenomena, identified as instantons, that can spoil the saddle 
point approximation, and have indicated the situation in which they do not. 
However, we have also shown that the instantons can cause $G(z)$ become 
undefined in certain regions of the complex $z$-plane. 

As examples, we have applied the introduced machinery to the Lego discrepancy 
and the $L_2^*$-discrepancy in one dimension. We identified the gauge freedom, 
calculated the zeroth order term of the expansion and gave $\prop$ in 
certain gauges. The zeroth order terms, which give $G_0$, are in agreement 
with with earlier calculations. 
We have shown that instantons appear for both of 
the discrepancies and that they are no threat for the perturbation series, but 
cause $G(z)$ to be undefined for asymptotically large $N$ when the real part 
of $z$ is 
larger then a certain positive value. For the $L_2^*$-case this value 
is $\half\pi^2$, the smallest positive value of $z$ at which $G_0$ has a 
singularity. 

Results of perturbative calculations supported by the techniques put forward 
in this paper will be presented in \cite{hak2}. 

\section*{Appendix A}
\addcontentsline{toc}{section}{Appendix A}
A more rigorous proof of \eqn{genG} for the Lego problem class goes as follows. 
According to \eqn{legodiscr}, the discrepancy is given by 
\begin{equation}
   D_N
   \;=\; \frac{1}{N}\sumM\frac{S_n^2}{w_n} - N \;\;,
\end{equation}
where $S_n=\sumN\vt_n(x_k)$ counts the number of points $x_k$ in bin $n$. 
If the points $x_k$ are truly randomly distributed, the variables $S_n$ are 
distributed according to a multinomial distribution so that the generating 
function can be written as 
\begin{equation}
   \Exp{e^{zD_N}}
   \;=\; \sum_{\{S_n\}}\frac{N!}{S_1!\cdots S_M!}\,
         w_1^{S_1}\cdots w_M^{S_M}\,
	 \exp\left(\frac{z}{N}\sumM\frac{S_n^2}{w_n} - zN\right) \;\;,
\label{multinom}	 
\end{equation}
where the summation is over all configurations $\{S_n\}$ which satisfy 
$\sumM S_n=N$. Notice that 
$\Exp{e^{zD_N}}>w_n^N\exp(zN/w_n-zN)$ 
for every $n$, so that the generating function is not defined if $N\ra\infty$ 
for the values of $z$ with $\textrm{Re}~z>\frac{w_n}{w_n-1}\log w_n$. 
Using Gaussian integration rules and the generalized binomial 
theorem, it is easy to see that \eqn{multinom} can be written as 
\begin{equation}
   \Exp{e^{zD_N}}
   \;=\; e^{-zN}\left(\prodM\frac{w_n}{2\pi}\right)^{\frac{1}{2}}
         \int_{\Real^M} \exp\left(-\frac{1}{2}\sumM w_ny_n^2\right)
	      \left(\sumM w_ne^{gy_n}\right)^N d^M\!y \;\;,        
\end{equation}
with $g=\sqrt{2z/N}$.
By writing the $N$-th power as a power of $e$ and substituting 
$y_n=\phi_n+Ng$, \eqn{Legogen} is obtained.

For the Wiener problem class, we can show that there is a na\"{\i}ve 
continuum limit which results in \eqn{WienG}. We use the fact that the 
discrepancy can be defined as the na\"{\i}ve continuum limit of 
\begin{align}
  &D_N^{(M)} 
   \;=\; \frac{1}{N}\sum_{\rho=1}^M\si_\rho^2
         \left(\sumN\sumM 
	 K_n^\rho\left[\vt_n(x_k)-w_n\right]\right)^2 \;\;, \label{gendiscr}\\
  &\vt_n(x) = \theta(\sfrac{n-1}{M}\leq x < \sfrac{n}{M}) \;,\;\;
   w_n = \int_0^1\vt_n(x)\,dx \;,\;\;
   K_n^\rho = \theta(\rho\leq n) \;,\;\;
   \si^2_\rho = \frac{1}{M} \;.
\end{align}
$D_N^{(M)}$ is the discretized version of the $L_2^*$-discrepancy, obtained
when in \eqn{defL_2^*} the average over a finite number of points $y_n$, 
$n=1,\ldots,M$ is taken, instead of the average over the whole of $\Kube$. 
Notice that a whole class of `discrete' discrepancies can be written as 
\eqn{gendiscr}, by choosing different expressions for the $K^\rho_n$ and the 
$\si^2_\rho$. Just like the Lego discrepancy, such a discrepancy can be written 
in terms of variables $S_n$ that count the number of points $x_k$ in bin
$n$, and is given by 
\begin{align}
  &D_N^{(M)}
   \;=\; \frac{1}{N}\dsumM R_{nm}S_nS_m - 2\sumM T_nS_n + NU \;\;,\notag\\
  &R_{nm} = \sum_{\rho=1}^M \si^2_\rho K^\rho_n K^\rho_m \;\;,\;\;
   T_n = \sum_{m=1}^M R_{nm}w_m \;\;,\;\;
   U = \dsumM R_{nm}w_nw_m \;\;.
\end{align}
In the case of the $L_2^*$-type discrepancy, the matrix $R$ is given by 
$R_{nm}=\min(n,m)/M$. The generating function is again given as the expectation 
value under the multinomial distribution. If we assume that the matrix $R$ is 
invertible and positive definite, as it is for the $L_2^*$-type discrepancy, 
use the Gaussian integration rules and the generalized binomial theorem and do 
the appropriate co-ordinate transformations, we find 
\begin{align}
  &G(z) 
  \;=\; \sqrt{\frac{\det R^{-1}}{(2\pi)^M}}\int_{\Real^M}
        \exp\left(-S[\phi]\right)d^M\!\phi \;\;, \notag\\
  &S[\phi] 
   \;=\;    \frac{1}{2}\dsumM R^{-1}_{nm}\phi_n\phi_m
          + \sqrt{2zN}\sumM w_n\phi_n
	  - N\log\left(\sumM w_ne^{\sqrt{\frac{2z}{N}}\,\phi_n}\right) \;\;.
\label{discWien1}	  
\end{align}
For the $L_2^*$-type discrepancy the inverse $R^{-1}$ of the matrix $R$ is easy 
to find and we get 
\begin{equation}
   \dsumM R^{-1}_{nm}\phi_n\phi_m
   \;=\; M\phi_1^2 + M\sum_{n=2}^M(\phi_n-\phi_{n-1})^2 \;\;,
\label{discWien2}   
\end{equation}
so that a na\"{\i}ve continuum limit clearly produces \eqn{WienG}. Notice that 
the term $M\phi_1^2$ gives the boundary condition $\phi(0)=0$; because $M$ 
becomes large, functions with $\phi(0)\neq0$ will not contribute to the path 
integral.

\section*{Appendix B: Matrices of the form 
                      $A_{nm}=a_n\de_{nm}+\epsilon\, b_nb_m$}
\addcontentsline{toc}{section}{Appendix B}
The eigenvalues $\la$ of a real-valued matrix $A$ are given by the zeros of the 
characteristic polynomial $P_A$. If $A$ is an $M\times M$ matrix with
matrix elements
\begin{equation}
   A_{nm} \;=\; a_n\de_{nm} +\epsilon\, b_nb_m \;\;,\quad 
   a_n,b_n\in\Real\;,\;\;n=1,\ldots,M\;\;,\quad\epsilon=\pm 1\;\;,
\label{defmat}   
\end{equation}
then the characteristic polynomial $P_A$ is given by 
\begin{equation}
   P_A(x) \;=\; Q_A(x)\prod_{n=1}^M(a_n-x) \;\;,\quad
   Q_A(x) \;=\; 1+\epsilon\sum_{m=1}^M\frac{b_m^2}{a_m-x} \;\;.
\end{equation}
For simplicity, we assume that the coefficients $a_n$ are ordered such that 
$a_1\leq a_2\leq\cdots\leq a_M$.
If a number of $d_n$ coefficients $a_n$ take the same value, that is, if 
$a_n$ is $d_n$-fold degenerate, then $\la=a_n$ is a $(d_n-1)$-fold degenerate 
eigenvalue of $A$. The remaining eigenvalues are given by the zeros of the 
function $Q_A$.
Except of the poles at $x=a_n$, $n=1,\ldots,M$, this function is continuous 
and differentiable on the whole of $\Real$. Furthermore, the sign of the 
derivative is equal to $\epsilon$. This
means that for each zero $\la$ of $Q_A$ except one, there is an $n$, such 
that $a_n<\la<a_{n+m}$ for the nearest and non-equal neighbor $a_{n+m}$ of 
$a_n$. The one other zero is smaller than $a_1$ if $\epsilon=-1$, 
and larger than $a_M$ if $\epsilon=1$. This 
is easy to see because $\lim_{x\ra\infty}Q_A(x)=\lim_{x\ra-\infty}Q_A(x)=1$.

\section*{Appendix C: Derivation of \eqn{diffTT1}}
\addcontentsline{toc}{section}{Appendix C}
We use the definitions of $T(E)$ as the r.h.s. of \eqn{period} and $T_1(E)$ 
as given in \eqn{defT1}:
\begin{equation}
   T(E) 
   \,=\, \int\limits_{\phi_-}^{\phi_+}\frac{d\phi}{\sqrt{E-U(\phi)}} \;,\quad
   T_1(E) 
   \,=\, \int\limits_{\phi_-}^{\phi_+}\frac{\phi\,d\phi}{\sqrt{E-U(\phi)}} 
   \;,\quad
   U(\phi) 
   = e^\phi - \phi - 1 \;.
\end{equation}
Because the end points $\phi_+$ and $\phi_-$ depend on $E$ such that 
$E-U(\phi_\pm)=0$, we can use Leibnitz's rule for differentiation under the 
integral sign to write
\begin{equation}
   T(E) \;=\; 
   2\,\frac{d}{dE}\int\limits_{\phi_-}^{\phi_+}\sqrt{E-U(\phi)}\,d\phi \;\;.
\end{equation}
Now we write $\sqrt{E-U(\phi)}=(E-e^\phi+\phi+1)(E-U(\phi))^{-1/2}$ and use 
that $1-e^\phi=dU/d\phi$, so that 
\begin{equation}
   T(E) \;=\; 
   2\,\frac{d}{dE}\left(
                   ET(E) + T_1(E) - \int\limits_{\phi_-}^{\phi_+}
                   \frac{1}{\sqrt{E-U(\phi)}}\frac{dU}{d\phi}\,d\phi\right)\;\;.
\end{equation}
But the last integral is equal to zero, and as a result, we obtain 
\eqn{diffTT1}.

\end{document}